\documentclass[aps,nofootinbib,showpacs,preprintnumbers,amsmath,amssymb]{revtex4}
\def\bes{\begin{subequations}}
\def\ees{\end{subequations}}
\def\be{\begin{equation}}
\def\ee{\end{equation}}
\def\bea{\begin{eqnarray}}
\def\eea{\end{eqnarray}}
\def\ba{\begin{eqnarray}}
\def\ea{\end{eqnarray}}
\def\bear{\begin{array}}
\def\eear{\end{array}}
\def\p1sl{\displaystyle{\not}p_1}
\def\p2sl{\displaystyle{\not}p_2}

\newcommand{\B}{{\overline B}}
\newcommand{\K}{{\widetilde {\cal K}}}
\newcommand{\G}{{\widetilde {\Gamma}}}
\newcommand{\bG}{{\overline {\Gamma}}}

\usepackage{slashed}
\usepackage{graphics}
\usepackage{graphicx}
\usepackage{dcolumn}
\usepackage{bm}
\usepackage{epsfig}
\usepackage{graphicx}
\usepackage{multirow}
\usepackage{dcolumn}
\usepackage{graphicx,epsfig}%

\begin{document}
\preprint{USM-TH-335}

\title{Oscillation of heavy sterile neutrino in decay of $B \to \mu e \pi$}
\author{Gorazd Cveti\v{c}$^1$}
\email{gorazd.cvetic@usm.cl}
\author{C.~S.~Kim$^2$}
\email{cskim@yonsei.ac.kr}
\author{Reinhart K\"ogerler$^3$}
\email{koeg@physik.uni-bielefeld.de}
\author{Jilberto Zamora-Sa\'a$^1$}
\email{jilberto.zamora@usm.cl}

\affiliation{$^1$
Department of Physics, Universidad T\'ecnica Federico Santa Mar\'ia, Casilla 110-V, Valpara\'iso, Chile\\
$^2$Department of Physics and IPAP, Yonsei University, Seoul 120-749, Korea\\
$^3$Department of Physics, Universit\"at Bielefled, 33501 Bielefeld, Germany}

\date{\today}

\begin{abstract}
In a scenario with two almost mass degenerate heavy sterile 
Majorana neutrinos with a
mass $\sim 1$ GeV, we present the semileptonic decay widths of heavy
charged pseudoscalars such as $B$ mesons, either 
lepton-number-violating ($B^{\pm} \to \mu^{\pm} e^{\pm} \pi^{\mp}$), 
or lepton-number-conserving ($B^{\pm} \to \mu^{\pm} e^{\mp} \pi^{\pm}$), 
mediated by such on-shell neutrinos.
It had been explained in the literature that such decays may be 
eventually detected, and that they can present even 
detectable CP violation effects. 
In this work we point out that, in addition, such decays may present
detectable effects of heavy neutrino oscillation, allowing us to
extract the oscillation length and thus the heavy neutrino mass difference
$\Delta M_N$, as well as a CP-violating Majorana phase.
\end{abstract}


\maketitle

\section{Introduction}

The neutrinos can be either Majorana or Dirac particles, although most of the 
neutrino scenarios suggest that the neutrinos are Majorana particles. In this
work we will assume that they are Majorana. As a consequence, they
can induce not just lepton number conserving (LNC) but also
lepton number violating (LNV) processes. Such processes are
neutrinoless double beta decays in nuclei \cite{0nubb},
specific scattering processes \cite{scatt1,scatt234,scatt3,scattDev,scattDas}
and rare meson decays \cite{RMDs,Atre,Boya,CDKK,CDK,CKZ,CKZ2,DCK,CDKZ}.

For neutrinos with masses, neutrino oscillations were predicted 
some time ago \cite{Pontecorvo}.  Oscillations of active (light)
neutrinos were later observed \cite{oscatm,oscsol,oscnuc}, with the
conclusion that the first three neutrinos have nonzero but
light masses $\alt 1$ eV. The oscillations are sensitive only to
mass differences, while neutrinoless double beta decays 
and rare meson decays
can help with the determination of the absolute mass of the light Majorana 
neutrinos. The best present upper bounds on the absolute masses of the light 
neutrinos are obtained from cosmology $m_\nu \gtrsim 0.23$ eV \cite{PlanckColl}.

The light neutrino masses can be produced via the seesaw mechanism \cite{seesaw}
where more than three neutrino flavors are required and where all of them are
Majorana. The light neutrinos in these seesaw scenarios have
masses $\sim {\cal M}_D^2/{\cal M}_R$ ($\alt 1$ eV),
with ${\cal M}_D$ being an electroweak scale or lower;
the heavy neutrinos are very heavy, with masses ${\cal M}_R \gg 1$ TeV,
their mixing with active neutrino flavors being very
suppressed $\sim {\cal M}_D/{\cal M}_R$ ($\ll 1$).
Other seesaw scenarios exist where the heavy neutrinos have lower masses
$M_N \lesssim 1$ TeV, Refs.~\cite{WWMMD},
and even $M_N \lesssim 1$ GeV \cite{scatt234,nuMSM,HeAAS,KS,AMP,NSZ};
their mixing with the standard model flavors may be less suppressed than 
in the original scenarios.

CP violation in the neutrino sector
in scenarios with nearly degenerate heavy neutrino masses 
had been investigated  in scattering processes in the
literature \cite{Pilaftsis} (resonant CP violation), as an effect
coming from the interference of tree-level with one-loop effects from the
neutrino propagators. Further, CP violation in the leptonic \cite{CKZ,CDKZ}
and semileptonic meson decays \cite{CKZ2,DCK,CDKZ} was investigated 
with a simpler, effectively tree-level, formalism and again in scenarios
with nearly degenerate on-shell heavy neutrino masses, and where the
decay width matrix of the massive neutrinos was assumed to be diagonal.

Scenarios with nearly degenerate heavy neutrino masses 
(and CP violation effects) appear in various
models, in particular in the neutrino minimal standard model ($\nu$MSM) 
\cite{nuMSM,Shapo} where these neutrinos generate baryon asymmetry of the
Universe while an additional lighter neutrino (with mass $\sim 10^1$ keV)
is responsible for the dark matter. Some general frameworks of low-scale seesaw
\cite{lsseesaw1,lsseesaw2} with more than two heavy neutrinos can also explain
the baryon asymmetry (but not the dark matter) and can have
larger values of the heavy-light mixing than in $\nu$MSM. In such 
models the almost mass degeneracy of Majorana neutrinos 
is preferred in the sense that it allows larger heavy-light mixings.

In this work we discuss neutrino oscillations in semileptonic decays of
heavy pseudoscalar mesons (such as $B$, $B_c$, $D_s$)
mediated by two on-shell Majorana neutrinos
$N_j$ ($j=1,2$) which are almost mass degenerate. Similar effects have
been investigated recently in leptonic decays of such mesons,
in Ref.~\cite{Boya}, where a quantum field theoretical generalization
of the Wigner-Weisskopf approach \cite{Boya2}
was implemented and used. Our approach is simpler, and the results
obtained are hopefully easier to interpret.

In Sec.~\ref{sec:gamma} we present the results for the decay widths
$\Gamma$ and the effective decay widths $\Gamma_{\rm eff}$ for the 
considered meson LNV semileptonic decay processes
$B^{\pm} \to \mu^{\pm} N_j \to \mu^{\pm} e^{\pm} \pi^{\mp}$
and their LNC counterparts
$B^{\pm} \to \mu^{\pm} N_j \to \mu^{\pm} e^{\mp} \pi^{\pm}$,
in the scenario with two on-shell almost mass degenerate
heavy neutrinos $N_j$, all without the neutrino oscillation effects.
Further, we also include the differential effective decay width
$d \Gamma_{\rm eff}(L)/dL$ for such processes, where $L$ is the
distance between the production ($\nu$-$N_j$) vertex and the
decay vertex ($N_j$-$e$-$\pi$). In Sec.~\ref{sec:osc} we then
extend these expressions by including the oscillation effects
of the on-shell neutrinos. In Appendix \ref{app:wf} we
show the consistency of the oscillation amplitude method applied in
Sec.~\ref{sec:osc} with the more usual quantum mechanics 
approach to oscillations. In Sec.~\ref{sec:est} we then estimate
numerically the oscillation length and describe the conditions under
which the oscillation modulation of the differential effective
decay width $d \Gamma_{\rm eff}(L)/dL$ can be measured.
In Sec.~\ref{sec:B} we indicate how the magnitudes
$|B_{\ell N_j}|$ of the heavy-light mixing parameters may be determined
by such measurements.
In Sec.~\ref{sec:concl} we summarize our results.

\label{intr}

\section{The decay width expression}
\label{sec:gamma}

In this section we present formulas for the LNV and LNC semileptonic decays
of charged $B$ mesons, of the type $B \to \mu e \pi$, mediated by
heavy sterile on-shell neutrinos. 
The formulas for the decay width of the 
LNV decays of this type,  $B^{\pm} \to \mu^{\pm} e^{\pm} \pi^{\mp}$,
in the case one heavy neutrino, were presented in Ref.~\cite{CDKK}.
They were extended to the case of two almost degenerate heavy on-shell
neutrinos in Ref.~\cite{CKZ2} (see also Ref.~\cite{DCK}), in the context
of CP violation. For a review we refer to \cite{CDKZ}.
We will use these formulas, and also the formulas
for the decay width of the LNC decays of this type,
$B^{\pm} \to \mu^{\pm} e^{\mp} \pi^{\pm}$.

There exist various scenarios with sterile neutrinos. 
Of particular interest is the 
$\nu$MSM of Shaposhnikov {\it et al.}, Refs.~\cite{nuMSM,Shapo}.
This model contains two almost degenerate Majorana neutrinos 
$N_j$ ($j=1,2$) of mass $\sim 1$ GeV and another lighter neutrino 
$\nu_K$ of mass $\sim 10^1$ keV,
as well as the three light neutrinos $\nu_j$ of mass  $\lesssim 1$ eV.
The striking advantage of this model is that it can explain
simultaneously the existence of neutrino
oscillations, dark matter and baryon asymmetry of the Universe.
We also wish to point out that
there exist more general (less constrained)
frameworks of the low-scale seesaw, which explain baryon 
asymmetry but not the dark matter. In such models,
larger values of the heavy-light mixing \cite{lsseesaw1} are allowed
than in $\nu$MSM, and the case of almost mass degeneracy 
of Majorana neutrinos is preferred \cite{lsseesaw2} 
since it allows larger mixings.

\begin{figure}[htb]
\centering\includegraphics[width=100mm]{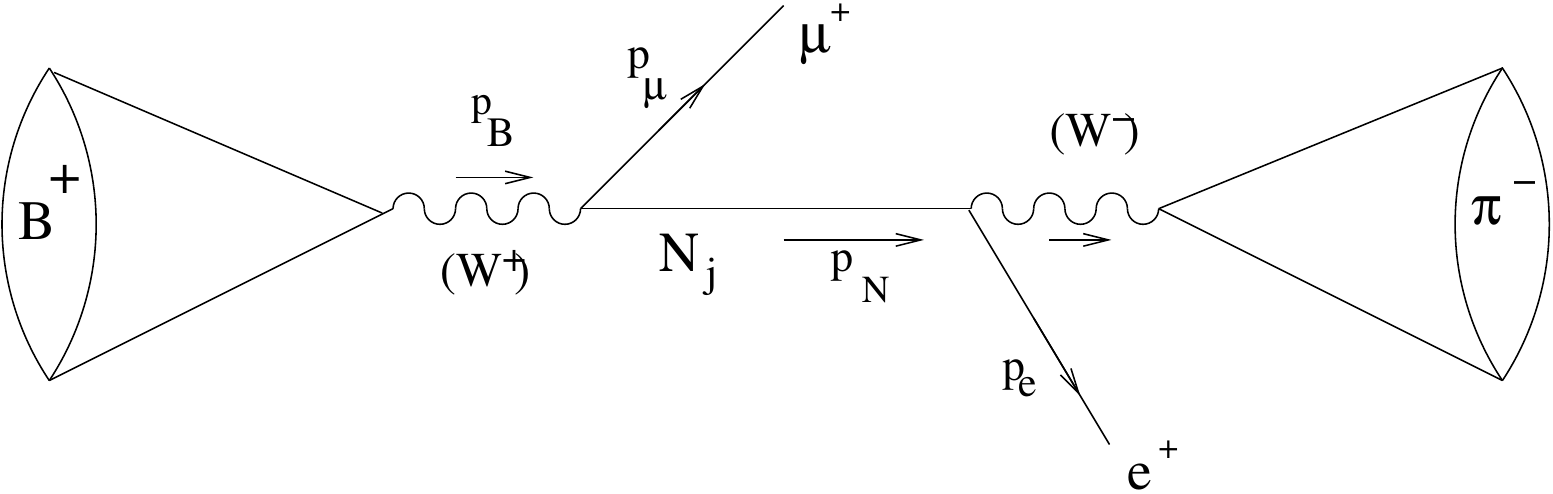}
\caption{The LNV decay $B^+ \to \mu^+ e^+ \pi^-$ 
via exchange of an on-shell neutrino $N_j$ ($j=1,2$).}
\label{FigBmuepiLNV}
\end{figure}

\begin{figure}[htb]
\centering\includegraphics[width=100mm]{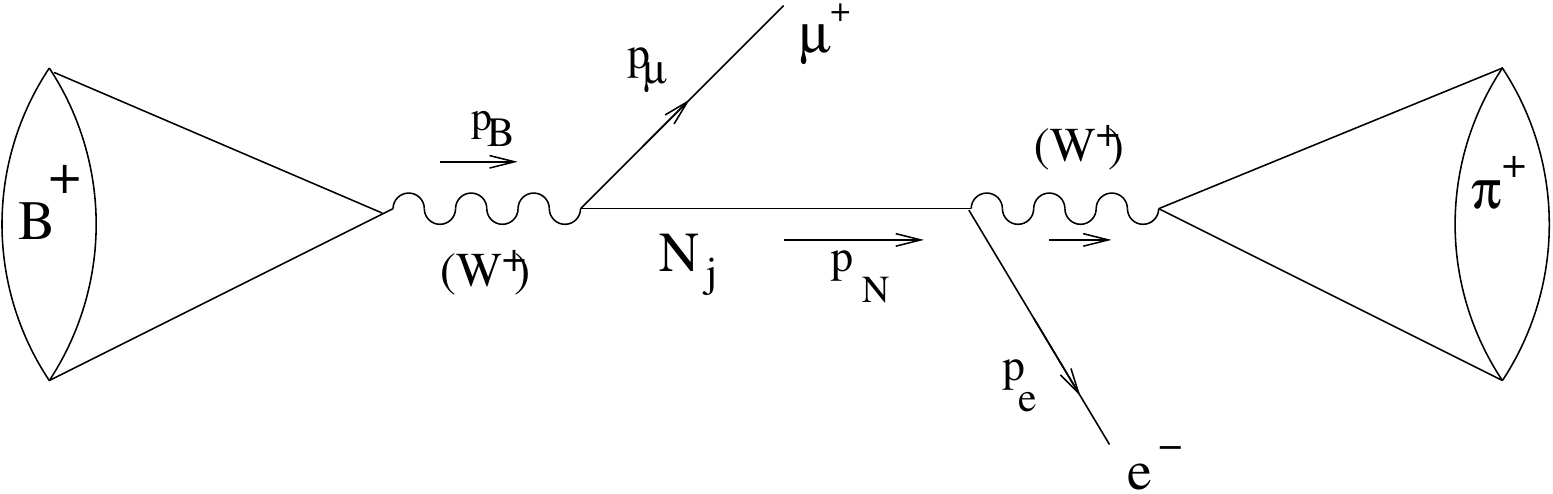}
\caption{The LNC decay $B^+ \to \mu^+ e^- \pi^+$ 
via exchange of an on-shell neutrino $N_j$ ($j=1,2$).}
\label{FigBmuepiLNC}
\end{figure}

The two flavor neutrinos $\nu_{e}$ and $\nu_{\mu}$ ($\ell =e, \mu, \tau$)
can be represented as
\bes
\label{mix}
\bea
\nu_{e} &=& \sum_{j=1}^3 B_{e j} \nu_j + B_{e N_1} N_1 + B_{e N_2} N_2 + \ldots,
\label{mixa}
\\
\nu_{\mu} &=& \sum_{j=1}^3 B_{\mu j} \nu_j + B_{\mu N_1} N_1 + B_{\mu N_2} N_2 +
\ldots.
\label{mixb}
\eea
\ees
The coupling of these two flavor neutrinos to the corresponding
charged leptons $e^{\pm}$ and $\mu^{\pm}$ has a part which contains
the coupling to the heavy almost mass-degenerate neutrinos $N_1$ and $N_2$
\bes
\label{ellWN}
\bea
{\cal L}_{e W N} &=& \frac{g}{2 \sqrt{2}} \left[
{\overline \psi}_{(e)} {\slashed{W}}_L \left(B_{e N_1} N_1 + B_{e N_2} N_2
\right) + {\rm h.c.} \right] =
K_1 \frac{g}{2 \sqrt{2}} {\overline \psi}_{(e)} {\slashed{W}}_L
{\cal N}_1 + {\rm h.c.}
\label{eWN}
\\
{\cal L}_{\mu W N} &=& \frac{g}{2 \sqrt{2}} \left[
{\overline \psi}_{(\mu)} {\slashed{W}}_L \left(B_{\mu N_1} N_1 + B_{\mu N_2} N_2
\right) + {\rm h.c.} \right] =
K_2 \frac{g}{2 \sqrt{2}} {\overline \psi}_{(\mu)} {\slashed{W}}_L
{\cal N}_2 + {\rm h.c.},
\label{muWN}
\eea
\ees
where we denoted by ${\cal N}_1$ and ${\cal N}_2$ the $e$- and $\mu$-flavor
analogs of the heavy neutrino mass eigenfields $N_j$ ($j=1,2$)
\bes
\label{cNs}
\bea
{\cal N}_1 &=& \B_{11} N_1 + \B_{12} N_2 
\nonumber\\
{\cal N}_2 &=& \B_{21} N_1 + \B_{22} N_2
\label{cN1N2}
\\
\B_{\alpha k} &\equiv& \frac{1}{K_{\alpha}} B_{\alpha k}, \qquad
K_{\alpha}  \equiv  \sqrt{|B_{\alpha 1}|^2 + |B_{\alpha 2}|^2} \qquad
(\alpha, k = 1, 2),
\label{bB}
\eea
\ees
where in $B_{\alpha k}$ the coefficients $\alpha=1,2$ stand for $e, \mu$, 
respectively; and $k=1,2$ for $N_1, N_2$, respectively.
The considered mechanisms for the LNV decays 
$B^{\pm} \to \mu^{\pm} e^{\pm} \pi^{\mp}$,  with on-shell $N_j$'s ($j=1,2$),
are those in Fig.~\ref{FigBmuepiLNV};
for the LNC decays, 
$B^{\pm} \to \mu^{\pm} e^{\mp} \pi^{\pm}$
are those in Fig.~\ref{FigBmuepiLNC}.
Although the LNV decays have also the crossed channel (i.e., the ones where
the vertices of $\mu$ and $e$ are exchanged), we assume here that the
measurements can distinguish these two channels (since $\mu \not= e$), 
by reconstructing invariant masses from the detected final state particles.

For these processes, we will consider the scenarios where the two
heavy neutrinos $N_1$ and $N_2$ are almost mass-degenerate ($\Delta M_N \ll
M_N \equiv M_{N_1}$) and are on shell. We will 
consider only the neutrino couplings (\ref{ellWN}),
with no components of the other mass eigenfields 
(thus no light mass eigenfields $\nu_1$, $\nu_2$, $\nu_3$),
because we will assume that $N_1$ and $N_2$ are the only
neutrinos which are on shell in these processes. This is then
reflected in our definition of ``heavy'' flavor states ${\cal N}_j$,
Eq.~(\ref{cN1N2}). We stress
that neutrinos which are off shell in these processes give, in relative terms, 
completely negligible contributions and will thus be ignored.

In the more general case of the LNV decay 
$M^{\pm} \to \ell_1^{\pm} N \to \ell_1^{\pm} \ell_2^{\pm} M^{' \mp}$
($\ell_j=e, \mu, \tau$; $M$ and $M^{'}$ pseudoscalars),
with $\ell_1 \not= \ell_2$ and neutrino $N$ on shell, the
corresponding partial widths can be written as 
\be
\Gamma(M^{\pm} \to  \ell_1^{\pm} N \to \ell_1^{\pm} \ell_2^{\pm} M^{' \mp}) = 
|B_{\ell_1 N}|^2 |B_{\ell_2 N}|^2 \; \G \qquad (\ell_1 \not= \ell_2),
\label{GamN}
\ee
where
\ba
\G & = &
 \frac{K^2 M_M^5}{64 \pi^2} \frac{M_{N}}{\Gamma_{N} }
\; \lambda^{1/2}(1, y_N,y_{\ell_1})
\; \lambda^{1/2} \left( 1, \frac{y^{'}}{y_N},\frac{y_{\ell_2}}{y_N} \right)
Q(y_N; y_{\ell_1}, y_{\ell_2},y^{'}) \ ,
\label{GDDN}
\ea
and the notations used in Eq.~(\ref{GDDN}) are
\bes
\label{notGDDN}
\ba
K^2 & = & G_F^4 f_M^2 f_{M^{'}}^2 | V_{Q_u Q_d} V_{q_u q_d} |^2 \ ,
\label{K2}
\\
\lambda(y_1,y_2,y_3) & = & y_1^2 + y_2^2 + y_3^2 - 2 y_1 y_2 - 2 y_2 y_3 - 2 y_3 y_1 \ ,
\label{lambdaN}
\\
y_N &=& \frac{M_{N}^2}{M_M^2} \ , \quad
y_{\ell_s} =  \frac{M_{\ell_s}^2}{M_M^2} \ , \quad
y^{'} =\frac{M_{M'}^2}{M_M^2} \ ,  \quad (\ell_s=\ell_1, \ell_2) \ ,
\label{yNs}
\ea
\ees
and  the function $Q(y_N; y_{\ell_1}, y_{\ell_2},y^{'})$ is 
(cf.~Ref.~\cite{CKZ2})
\bes
\label{Q}
\bea
Q(y_N; y_{\ell_1}, y_{\ell_2}, y') & = &
{\bigg \{} \frac{1}{2}(y_N - y_{\ell_1})(y_N - y_{\ell_2})(1 - y_N - y_{\ell_1})
\left( 1 - \frac{y'}{y_N} +\frac{y_{\ell_2}}{y_N} \right)
\nonumber\\
&& +  {\big [}
- y_{\ell_1} y_{\ell_2} (1 + y' + 2 y_N - y_{\ell_1} - y_{\ell_2} )
- y_{\ell_1}^2 (y_N - y') + y_{\ell_2}^2 (1 - y_N)
\nonumber\\
&&
+ y_{\ell_1} (1+y_N) (y_N - y') - y_{\ell_2} (1-y_N)(y_N+y') {\big ]}
{\bigg \}} \ .
\label{Qa}
\\
& = & \frac{1}{2} \left[
(1 - y_N) y_N + y_{\ell_1} (1 + 2 y_{N} - y_{\ell_1}) \right] 
\left[y_N - y' - 2 y_{\ell_2} - \frac{y_{\ell_2}}{y_N} (y' - y_{\ell_2}) 
\right]
\label{Qb}
\eea
\ees
In Eq.~(\ref{K2}), $f_M$ and $f_{M^{'}}$ are the decay constants,
and $V_{Q_u Q_d}$ and $V_{q_u q_d}$ are the Cabibbo-Kobayashi-Maskawa
(CKM) matrix elements of pseudoscalars
$M^{\pm}$ and $M^{' \mp}$. 

We notice that in the considered specific case ($\ell_1=\mu$ and $\ell_2=e$)
we have $y_{\ell_1} \approx y_{\ell_2} \approx 0$, and expression (\ref{Q})
simplifies according to
\bea
 Q(y_N; y_{\ell_1}, y_{\ell_2}, y') \approx Q(y_N; 0, 0, y')
& = & \frac{1}{2} \left[ y_N (1 - y_N) \right] \left[ y_N - y' \right].
\label{Qyell0}
\eea

The results (\ref{GamN})-(\ref{Q}) can be written in an equivalent form
\bea
\Gamma(M^{\pm} \to \ell_1^{\pm} N \to \ell_1^{\pm} \ell_2^{\pm} M^{' \mp}) 
&=&
\frac{1}{\Gamma_N} \Gamma(M^{\pm} \to \ell_1^{\pm} N) 
\Gamma(N \to \ell_2^{\pm} M^{' \mp}) \qquad (\ell_1 \not= \ell_2),
\label{fact}
\eea
where the widths of the two decays are
\bes
\label{GbG}
\bea
 \Gamma(M^{\pm} \to \ell_1^{\pm} N) & = &
|B_{\ell_1 N}|^2 \bG(M^{\pm} \to \ell_1^{\pm} N),
\label{GbGa}
\\
 \Gamma(N \to \ell_2^{\pm} M^{' \mp}) & = &
|B_{\ell_2 N}|^2 \bG(N \to \ell_2^{\pm} M^{' \mp}),
\label{GbGb}
\eea
\ees
and the expressions for the corresponding canonical widths $\bG$ (i.e., widths
without the mixing factors) are
\bes
\label{bG}
\bea
\bG(M^{\pm} \to \ell_1^{\pm} N) & = &
\frac{1}{8 \pi} G_F^2 f_M^2 |V_{Q_u Q_d}|^2 M_M^3 
\; \lambda^{1/2}(1,y_N,y_{\ell_1}) 
 \left[ (1 - y_N) y_N + y_{\ell_1} (1 + 2 y_{N} - y_{\ell_1}) \right],
\label{bGM}
\\
\bG(N \to \ell_2^{\pm} M^{' \mp}) & = &
\frac{1}{16 \pi} G_F^2 f_{M^{'}}^2 |V_{q_u q_d}|^2 \frac{1}{M_N} 
\; \lambda^{1/2}\left( 1,\frac{y^{'}}{y_N},\frac{y_{\ell_2}}{y_N} \right) 
\left[ (M_N^2 + M_{\ell_2}^2)(M_N^2 - M_{M^{'}}^2+M_{\ell_2}^2) -
4 M_N^2 M_{\ell_2}^2 \right]
\nonumber\\
& = &
\frac{1}{16 \pi} G_F^2 f_{M^{'}}^2 |V_{q_u q_d}|^2 M_M^2 M_N 
\; \lambda^{1/2}\left( 1,\frac{y^{'}}{y_N},\frac{y_{\ell_2}}{y_N} \right)
\left[y_N - y' - 2 y_{\ell_2} - \frac{y_{\ell_2}}{y_N} (y' - y_{\ell_2}) 
\right],
\label{bGN}
\eea
\ees
where again the notations (\ref{notGDDN}) were used. We notice that
the algebraic factorization of the $Q$ function, Eq.~(\ref{Qb}), yields the
factorization (\ref{fact}), as can be seen by inspection of
the expressions (\ref{bGM}) and (\ref{bGN}).

It can be checked that the result for the LNC processes
$M^{\pm} \to \ell_1^{\pm} N \to \ell_1^{\pm} \ell_2^{\mp} M^{' \pm}$
is the same as the result (\ref{fact})-(\ref{GbG})
\bes
\label{factLNC}
\bea
\Gamma(M^{\pm} \to \ell_1^{\pm} N \to \ell_1^{\pm} \ell_2^{\mp} M^{' \pm}) 
&=&
\frac{1}{\Gamma_N} \Gamma(M^{\pm} \to \ell_1^{\pm} N) 
\Gamma(N \to \ell_2^{\mp} M^{' \pm})
\label{factLNCa}
\\ 
&=& \frac{|B_{\ell_1 N}|^2 |B_{\ell_2 N}|^2 }{\Gamma_N}
\bG(M^{+} \to \ell_1^{+} N) \bG(N \to \ell_2^{+} M^{' -})
\quad (\ell_1 \not= \ell_2),
\label{factLNCb}
\eea
\ees
where the canonical decay widths ($\bG$'s) are again those of Eq.~(\ref{bG}). 

\begin{figure}[htb] 
\begin{minipage}[b]{.49\linewidth}
\centering\includegraphics[width=85mm]{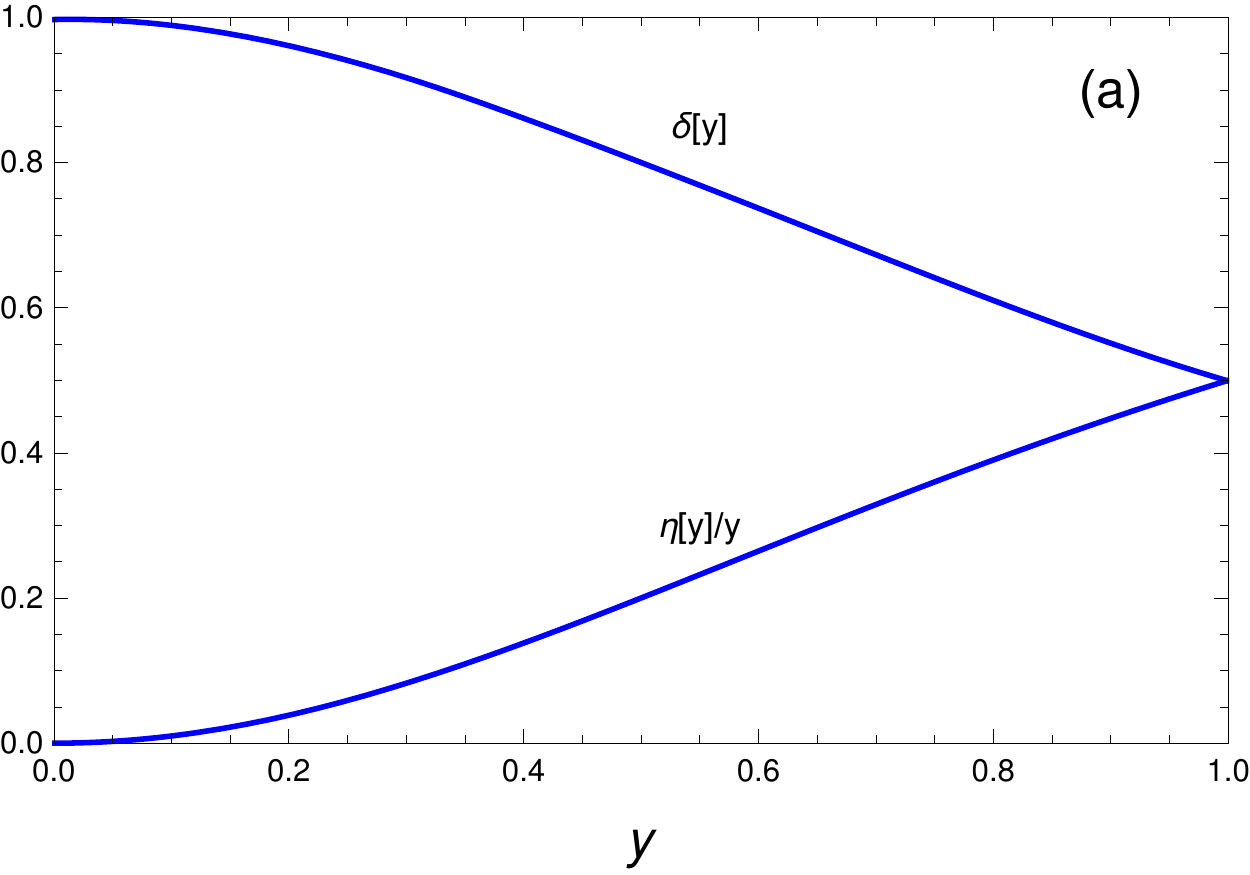}
\end{minipage}
\begin{minipage}[b]{.49\linewidth}
\centering\includegraphics[width=85mm]{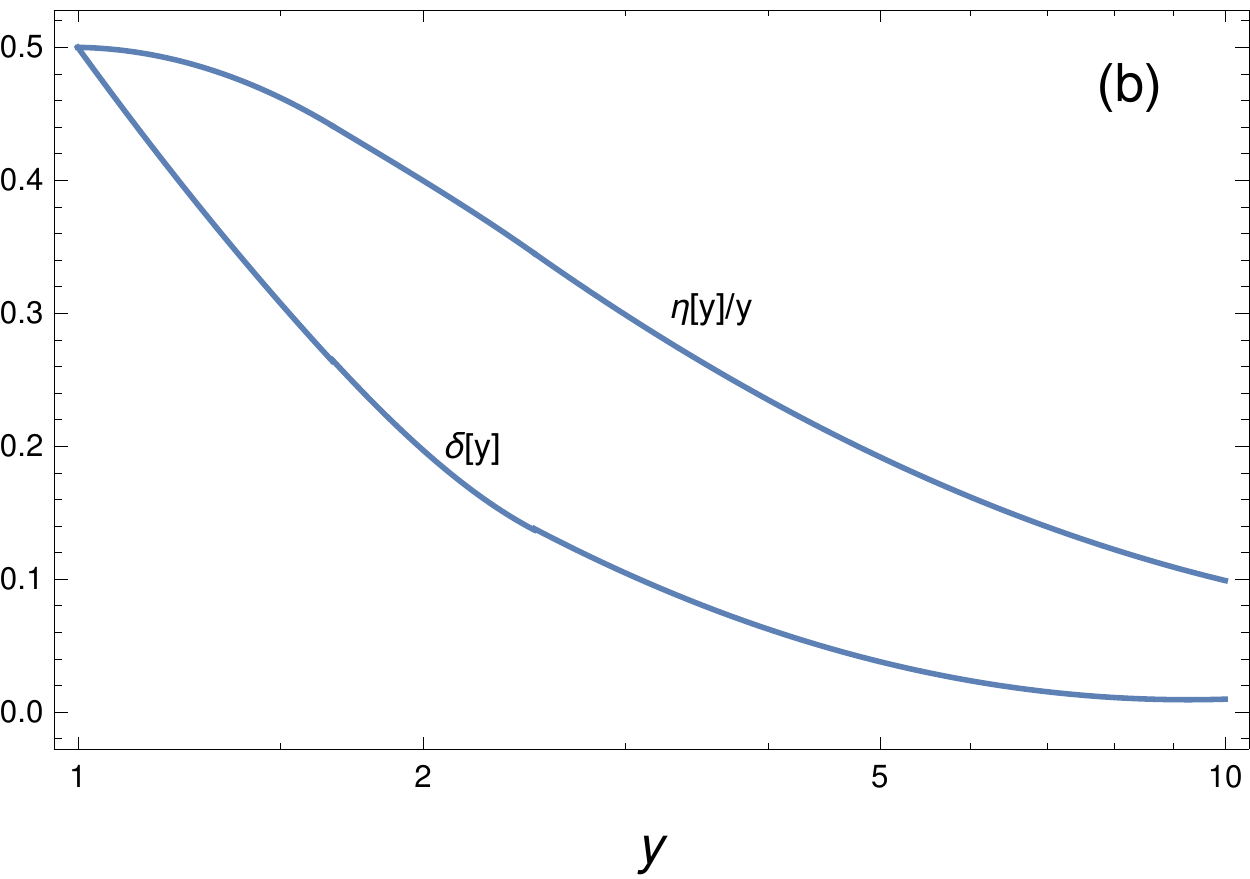}
\end{minipage}
\caption{The suppression factors $\eta(y)/y$ and $\delta(y)$
as a function of $y \equiv \Delta M_N/\Gamma_N$:
(a) for $0<y < 1$ (on the linear $y$ scale);
(b) for $1 < y < 10$ (on the logarithmic $y$ scale). Figures taken from
Refs.~\cite{CKZ2,CDKZ}. For $y<0$ we have $\delta(y)=\delta(-y)$ and
$\eta(y)=\eta(-y)$.}
\label{etadelfig}
\end{figure}

We recall that we will consider the scenario with two on-shell neutrinos
$N_j$ ($j=1,2$), and with almost degenerate masses: $|\Delta M_N| \ll M_{N_1}$,
where $\Delta M_N \equiv M_{N_2}-M_{N_1}$. In this case, it turns out that
the expression for the LNV decay width becomes more complicated,
cf.~Ref.~\cite{CKZ2}. With the notation (\ref{ellWN}) for the mixing
coefficients, it can be written in the following form:
\bea
\Gamma(B^{\pm} \to \mu^{\pm} e^{\pm} \pi^{\mp}) &=&  
\bG(B^+ \to \mu^+ N) \bG(N \to e^+ \pi^-)
\times {\big \{} \frac{|B_{\mu N_1}|^2 |B_{e N_1}|^2}{\Gamma_{N_1}} + 
+ \frac{|B_{\mu N_2}|^2 |B_{e N_2}|^2}{\Gamma_{N_2}}
\nonumber\\
&&
+  \frac{4}{(\Gamma_{N_1}+\Gamma_{N_2})} |B_{\mu N_1}| |B_{e N_1}| |B_{\mu N_2}| |B_{e N_2}| 
\left( \delta(y) \cos \theta_{21}^{\rm (LNV)}
\mp \frac{\eta(y)}{y} \sin \theta_{21}^{\rm (LNV)} \right)
{\big \}},
\label{GDDLNV}
\eea
where $M_N \equiv M_{N_1} \approx M_{N_2}$,
the angle $\theta_{21^{\rm (LNV)}}$ is a combination of the
phases of the heavy-light mixing coefficients
\be
\theta_{21}^{\rm (LNV)} = {\rm arg}(B_{\mu N_2}) + {\rm arg}(B_{e N_2}) -{\rm arg}(B_{\mu N_1}) - {\rm arg}(B_{e N_1}).
\label{theta21LNV}
\ee
The functions $\delta(y)$ and $\eta(y)/y$ appearing in Eq.~(\ref{GDDLNV})
are functions of the parameter $y \equiv \Delta M_N/\Gamma_N$ only,
where $\Gamma_N$ is the arithmetic average of the total decay widths
of $N_1$ and $N_2$\footnote{
For simplicity we assume that the $(2 \times 2)$ decay width matrix
$(\Gamma_N)_{ij}$ of the (near mass degenerate) neutrinos $N_1$ and $N_2$
has no (or negligible) off-diagonal element [$(\Gamma_N)_{12}$]. This
assumption was also taken in Refs.~\cite{CKZ,CKZ2,DCK,CDKZ}.}
\be
\Gamma_N = \frac{1}{2} \left( \Gamma_{N_1} + \Gamma_{N_2} \right), \qquad
y \equiv \frac{\Delta M_N}{\Gamma_N},
\label{GNy}
\ee
and the functions are presented in Fig.~\ref{etadelfig} for $y>0$.
It is straightforward to verify that the invariance of
expression (\ref{GDDLNV}) under the exchange of
the roles of $N_1$ and $N_2$ means that for $y<0$ we have
$\delta(y)=\delta(-y)$ and $\eta(y)=\eta(-y)$. 
The factors (functions) $\delta(y)$ and $\eta(y)/y$ represent the effects of the
$N_1$-$N_2$ overlap in the decay width, in the real and imaginary
parts of the $N_1$-$N_2$ interference terms, respectively.
Therefore, $\delta(y)$ and $\eta(y)/y$ go to zero
when $|y| \gg 1$, i.e., when no overlap. 
While both the $\delta(y)$ and the $\eta(y)/y$ function were obtained in
Ref.~\cite{CKZ2} numerically, it can be argued that $\eta(y)/y$ is
a simple function \cite{CDKZ}, $\eta(y)/y = y/(y^2+1)$,
and this agrees with the numerical results of Ref.~\cite{CKZ2}.
For details, we refer to Refs.~\cite{CKZ2,CDKZ}.

Finally, the expressions for the total decay widths 
$\Gamma_{N_1}$ and $\Gamma_{N_2}$ appearing in Eq.~(\ref{GDDLNV}) are 
\begin{equation}
\Gamma_{N_j} = \K_j \bG_N(M_{N}) \ ,
\label{GNwidth}
\end{equation}
where we denote $M_{N} = M_{N_1} \approx M_{N_2}$ and 
\begin{equation}
 \bG_N(M_{N}) \equiv \frac{G_F^2 M_{N}^5}{96 \pi^3} \ ,
\label{barGN}
\ee
and the factors $\K_j \sim |B_{\ell N_j}|^2$ ($j=1,2$) 
contain all the dependence on the heavy-light mixing factors
\begin{equation}
\K_j = {\cal N}_{e N} \; |B_{e N_j}|^2 + {\cal N}_{\mu N} \; |B_{\mu N_j}|^2 + {\cal N}_{\tau N} \; |B_{\tau N_j}|^2  \ .
\label{calK}
\end{equation}
In this expression, the coefficient functions
${\cal N}_{\ell N}(M_N) \equiv {\cal N}_{\ell N}$ ($\ell = e, \mu, \tau$)
depend only on the mass $M_N$ of the neutrino $N_j$;
these coefficient functions are $\sim 10^0$-$10^1$.
The curves for ${\cal N}_{\ell N}(M_N)$
as a function of $M_N$ were presented
in Ref.~\cite{CKZ2} for the case of Majorana neutrinos,
and in Ref.~\cite{CDKZ} for both cases of Dirac and Majorana neutrinos,
in the neutrino mass interval $0.1 \ {\rm GeV} < M_N < 6.3 \ {\rm GeV}$.

As mentioned earlier, in addition to the above LNV decay width,
there exists also the LNC decay width 
$\Gamma(B^{\pm} \to \mu^{\pm} e^{\mp} \pi^{\pm}$),
which in the case of scenario of one on-shell neutrino $N$ coincides with
the LNV expression (\ref{GDDN}). In the scenario with two on-shell
almost degenerate neutrinos $N_j$, the expression is slightly different
from the LNV equation (\ref{GDDLNV}); namely, only the angle 
$\theta_{21}^{\rm (LNV)}$ [Eq.~(\ref{theta21LNV})]
is now replaced by the following angle:
\be
\theta_{21}^{\rm (LNC)} = {\rm arg}(B_{\mu N_2}) - {\rm arg}(B_{e N_2}) -{\rm arg}(B_{\mu N_1}) + {\rm arg}(B_{e N_1}).
\label{theta21LNC}
\ee

We will consider, from now on,
the case when there is an almost degeneracy of the two heavy
neutrinos ($|\Delta M_N| \ll M_N \equiv M_{N_1}$) and at the same time 
the degeneracy $|\Delta M_N|$ is significantly larger than the (extremely
small) decay width $\Gamma_N$: 
\bea
|\Delta M_N| &\ll& M_N \; {\rm and} \; 
|y| \equiv \frac{|\Delta M_N|}{\Gamma_N} \gg 1.
\label{cond}
\eea
In this case, we can see from Fig.~\ref{etadelfig} that the functions
$\delta(y)$ and $\eta(y)/y$ become very small. Therefore,
the $N_1$-$N_2$ overlap term in $\Gamma(B^{\pm} \to \mu^{\pm} e^{\pm} \pi^{\mp})$
becomes negligible. As a result, 
Eq.~(\ref{GDDLNV}) reduces to the following form:
\bes
\label{GDDLNV2}
\bea
\Gamma(B^{\pm} \to \mu^{\pm} e^{\pm} \pi^{\mp}) &\approx&
\sum_{j=1}^2 
\Gamma(B^{\pm} \to \mu^{\pm} N_j \to \mu^{\pm} e^{\pm} \pi^{\mp}) 
\label{GDDLNV2a}
\\
&=&
\bG(B^{+} \to \mu^{+} N) \bG(N \to e^{+} \pi^{-})
\left\{ \frac{1}{\Gamma_{N_1}}|B_{\mu N_1}|^2 |B_{e N_1}|^2  +
\frac{1}{\Gamma_{N_2}} |B_{\mu N_2}|^2 |B_{e N_2}|^2 \right\}.
\label{GDDLNV2b}
\eea
\ees

The decay width presented hitherto does not contain an important
suppression (acceptance) factor. Namely, the on-shell neutrino
$N_j$ travels before decaying. The decay will be detected if the
on-shell neutrino decays during the passage of the neutrino through
the detector. If the length of the detector is $L$, then the probability
$P_N$ of decay of $N$ there is
\bes
\label{PN}
\bea
P_N(L) = 1 - \exp\left( - \frac{t}{\tau_N \gamma_N} \right)
&=& 1  - \exp\left( - \frac{L}{\tau_N \gamma_N \beta_N} \right)
\label{PNa}
\\
\approx  L/(\tau_N \gamma_N \beta_N) \qquad {\rm if} \; P_N\ll 1.
\label{PNb}
\eea
\ees
In the second identity of Eq.~(\ref{PNa}) we took into account that
$L = \beta_N t$ where $\beta_N$ ($\lesssim 1$)
is the velocity of the $N$ neutrino in the lab frame.
Furthermore, $\gamma_N = (1- \beta_N^2)^{-1/2}$ is the  Lorentz lab time dilation
factor, typically $\gamma_N > 2$. The $N$ lifetime in the rest is
$\tau_N = 1/\Gamma_N$. Therefore, $\gamma_N \tau_N$ is the $N$ lifetime 
in the lab frame.
The formula (\ref{PNb}) holds if $P_N \ll 1$, and we will
assume this to be the case in the considered cases. 

This decay-within-the-detector
probability $P_N$ has been discussed in 
Refs.~\cite{CDK,scatt3,CKZ,CKZ2,CERN-SPS,Gronau,commKim,CDKZ}
It is convenient to define the corresponding canonical, independent of mixing,
probability ${\overline P}_N$
\bes
\label{bPN}
\bea
{\overline P}_N(L) &=&  1 {\rm m} \times \frac{\bG(M_N)} {\gamma_N}  
\label{bPNa}
\\
\Rightarrow
P_N(L) &\approx & \left( \frac{L}{1 {\rm m}} \right) \times {\overline P}_N  \K.
\label{bPNb}
\eea
\ees
The quantity ${\overline P}_N$ is presented, 
for $\gamma_N =2$, in Fig.~\ref{bPNfig}
as a function of $M_N$.
\begin{figure}[htb]
\centering\includegraphics[width=120mm]{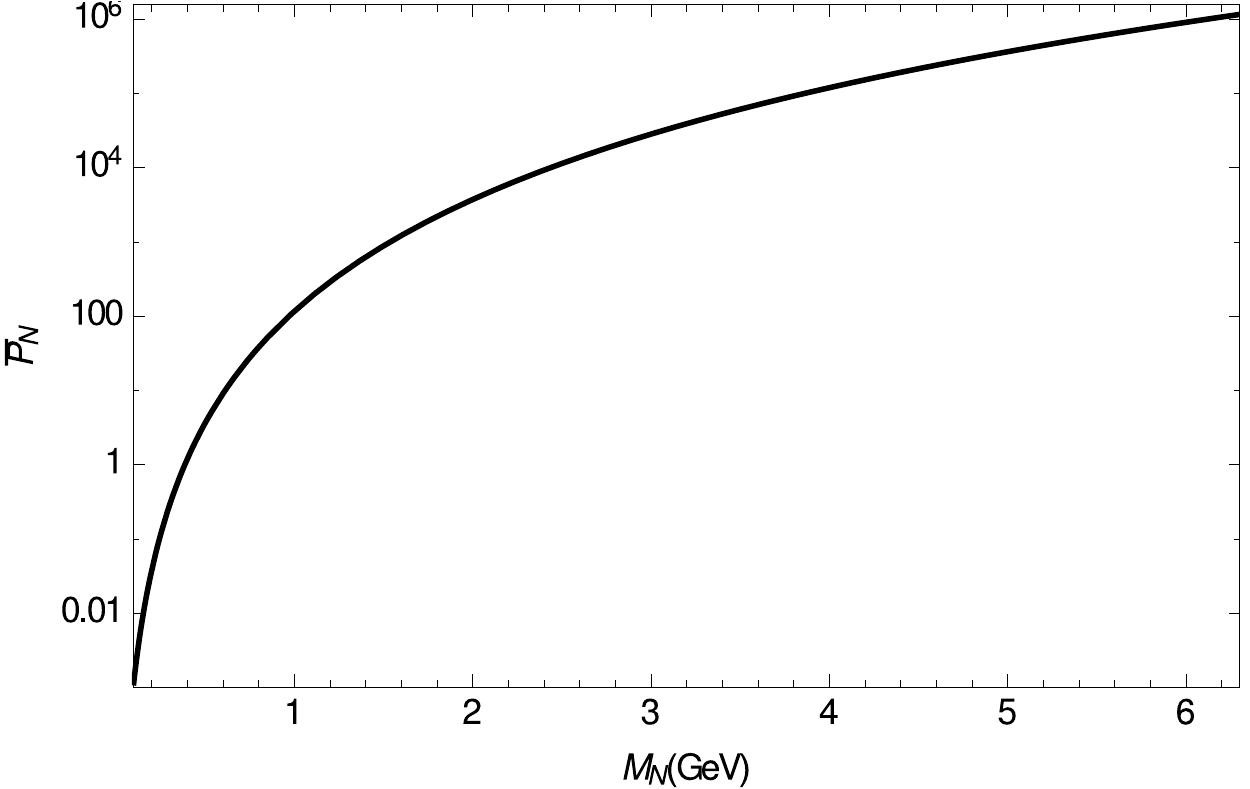}
\caption{The canonical probability ${\overline P}_N$,
as defined in Eq.~(\ref{bPN}), 
as a function of the neutrino mass $M_N$, with the Lorentz lab time dilation
factor chosen to be $\gamma_{N}$ [$\equiv (1 - \beta_N^2)^{-1/2}$] $=2$.}
\label{bPNfig}
\end{figure}

The effective (true) decay widths and branching ratios are
those multiplied by $P_N$. However, since we have two different 
(but almost mass degenerate) neutrinos $N_j$,
we have for each of them a different decay probability 
\be
P_{N_j}(L) \approx  
\left( \frac{L}{1 {\rm m}} \right) \times {\overline P}_N  \K_j
= \frac{L}{\gamma_N} \Gamma_{N_j}, 
\label{bPNj}
\ee
where $\K_j$ is given in Eq.~(\ref{calK}), and in the second equality
we used the relations (\ref{bPN}) and (\ref{GNwidth}). 
The canonical probability
${\overline P}_N$, Eq.~(\ref{bPN}), is common to both neutrinos $N_j$
because they have practically the same mass and thus the same kinematics
(and hence the same Lorentz factor $\gamma_N$).
The coefficients ${\cal N}_{\ell N}$($\sim 10^0$-$10^1$) in $\K_j$
are common to both neutrinos $N_j$ (because they have a practically equal mass);
but the mixings $B_{\ell N_j}$ can be, in principle, quite different for the
two neutrinos, and thus the two mixing factors $\K_j$ ($j=1,2$)
may differ significantly from each other.

Combining the probabilities (\ref{bPNj}) with the decay width (\ref{GDDLNV2b})
leads to the effective (true) decay width, where the dependence on 
the two decay widths $\Gamma_{N_j}$ cancels out:
\bes
\label{GLNVeff}
\bea
\Gamma_{\rm eff}(B^{\pm} \to \mu^{\pm} e^{\pm} \pi^{\mp};L) &\approx&
\sum_{j=1}^2 
\Gamma(B^{\pm} \to \mu^{\pm} N_j \to \mu^{\pm} e^{\pm} \pi^{\mp}) P_{N_j}(L)
\label{GLNVeffa}
\\
&\approx&
\frac{L}{\gamma_N \beta_N}
\bG(B^{+} \to \mu^{+} N) \bG(N \to e^{+} \pi^{-})
\left[|B_{\mu N_1}|^2 |B_{e N_1}|^2  + |B_{\mu N_2}|^2 |B_{e N_2}|^2 \right].
\label{GLNVeffb}
\eea
\ees
This implies that the effective differential decay, with respect to the
distance $L$ between the two vertices of the process, is
\bea
\frac{d}{d L}
\Gamma_{\rm eff}(B^{\pm} \to \mu^{\pm} e^{\pm} \pi^{\mp};L) &\approx&
\frac{1}{\gamma_N \beta_N}
\bG(B^{+} \to \mu^{+} N) \bG(N \to e^{+} \pi^{-})
\left[|B_{\mu N_1}|^2 |B_{e N_1}|^2  + |B_{\mu N_2}|^2 |B_{e N_2}|^2 \right],
\label{dGLNVeff}
\eea
which is independent of the distance $L$.

For the LNC processes $B^{\pm} \to \mu^{\pm} e^{\mp} \pi^{\pm}$,
the result is the same as in the above LNV processes, due to the
equality of the LNC decay width (\ref{factLNC}) with the LNV 
decay width in (\ref{fact}) and (\ref{GbG}) [cf.~also Eq.~(\ref{bG})].
Therefore, when $\delta(y), |\eta(y)/y| \ll 1$
[i.e., when the conditions (\ref{cond}) hold], we have
\bes
\label{LNCeff}
\bea
\frac{d}{d L}
\Gamma_{\rm eff}(B^{\pm} \to \mu^{\pm} e^{\mp} \pi^{\pm};L) & =&
\frac{d}{d L}
\Gamma_{\rm eff}(B^{\pm} \to \mu^{\pm} e^{\pm} \pi^{\mp};L),
\label{dGLNCeff}
\\
\Gamma_{\rm eff}(B^{\pm} \to \mu^{\pm} e^{\mp} \pi^{\pm};L) &=&
\Gamma_{\rm eff}(B^{\pm} \to \mu^{\pm} e^{\pm} \pi^{\mp};L).
\label{GLNCeff}
\eea
\ees

The present upper bounds for the
$|B_{\ell N_j}|^2$ mixing coefficients appearing in these expressions,
in the considered mass range $M_N \approx 1$-$5$ GeV, are
$|B_{\ell N_j}|^2 \sim 10^{-7}$-$10^{-4}$, cf.~\cite{Atre}.\footnote{
The present upper bounds for $|B_{\tau N}|^2$ are higher than that,
but they are expected to become significantly lower in the future.}

\section{The effects of neutrino oscillation}
\label{sec:osc}

In the previous section, important effects of neutrino
oscillation of the propagating on-shell neutrino were not accounted for.
As we will see, these effects lead to a modulation, i.e.,
the $L$-dependence of the effective decay widths obtained in the previous section,
where $L$ is the distance traveled by the on-shell neutrino between
its production and detection points ($L \approx \beta_N t$).

We will follow the lines of the approach of Ref.~\cite{Glash}
to neutrino oscillations. For the LNV
decays $B^+ \to \mu^+ N_j \to \mu^+ e^+ \pi^-$ of Fig.~\ref{FigBmuepiLNV},
the relevant interactions at the first (production) vertex are 
$- B_{\mu N_j}^* {\overline \mu^c} \gamma^{\eta} (1 + \gamma_5) N_j W_{\eta}^{(+)}$, 
and the neutrino state produced at this vertex is
\be
|\psi \rangle_{(B^+)} \sim 
B_{\mu N_1}^* | N_1(p_{N_1}) \rangle + B_{\mu N_2}^* | N_2(p_{N_2}) \rangle,
\label{psiBpl}
\ee
where the momenta of the two physical on-shell neutrinos
are slightly different from each other, because $|\Delta M_N| \not=0$
(we recall that $|\Delta M_N| \ll M_N$). We have 
\be
p_{N_j} = (E_{N_j},0,0,p_{N_j}^3), \qquad E_{N_j} = \sqrt{M_{N_j}^2 + (p_{N_j}^3)^2},
\label{pNj}
\ee
where the restriction to one spatial dimension (${\hat z}$) was made, because
the processes with oscillation require the neutrino to propagate far from
the production vertex. At the second vertex of the LNV process 
Fig.~\ref{FigBmuepiLNV}, the relevant coupling is 
$B_{e N_j}^* {\overline {N_j}} \gamma^{\delta} (1 - \gamma_5) e W_{\delta}^{(+)}$.
The detection of the neutrino there can be described by an operator at
the detector space-time location $z = (t, 0,0,L)$ where $L \approx \beta_N t$.
This operator is the annihilation operator
$B_{e N_j}^* {\hat b}_{(N_j)}(p_{N_j};z) =
B_{e N_j}^* {\hat b}_{(N_j)}(p_{N_j}) \exp(-i p_{N_j} \cdot z)$
acting at the aforementioned component $| N_j(p_{N_j}) \rangle
\sim {\hat b}_{(N_j)}(p_{N_j})^{\dagger} | 0 \rangle$ ($j=1,2$).
Since ${\hat b}_{(N)}(p_N) {\hat b}_{(N)}(p_N)^{\dagger} | 0 \rangle 
= {\rm const} | 0 \rangle$, this implies the following detection
amplitude\footnote{We use the metric $(1,-1,-1,-1)$ for the scalar products.
It is our understanding that the authors of Ref.~\cite{Glash} use the
metric $(-1,1,1,1)$.}:
\be
{\cal A}(B^+ \to \mu^+ e^+ \pi^{-}; L) \sim 
B_{\mu N_1}^* B_{e N_1}^* \exp(- i p_{N_1} \cdot z) +
B_{\mu N_2}^* B_{e N_2}^* \exp(- i p_{N_2} \cdot z).
\label{calAMpl}
\ee
The $L$ dependence of the effective (true) decay width of the
considered process is proportional to the absolute square of the
above amplitude 
\bes
\label{dGLNVeffo1}
\bea
\frac{d}{d L}
\Gamma_{\rm eff}^{\rm (osc)}(B^+ \to \mu^+ e^+ \pi^-;L) & \equiv &
\frac{1}{dL} \Gamma_{\rm eff}^{\rm (osc)}(B^+ \to \mu^+ e^+ \pi^-; L < L' < L+dL)
\sim  |{\cal A}(B^+ \to \mu^+ e^+ \pi^{-})|^2
\label{dGLNVeffo1a}
\\
& \sim & \left\{ \sum_{j=1}^2 |B_{\mu N_j}|^2 |B_{e N_j}|^2 +
2 {\rm Re} \left[ B_{\mu N_1}^*   B_{e N_1}^*   B_{\mu N_2} B_{e N_2}
\exp\left[ i (p_{N_2} - p_{N_1}) \cdot z \right] \right] \right\}.
\label{dGLNVeffo1b}
\eea
\ees
The superscript (osc) indicates that this is 
the (differential) effective decay width with oscillation effects included. 
The oscillation term, in comparison with expression (\ref{dGLNVeff}),
is new and introduces $L$-dependence in the otherwise $L$-independent
differential decay width $d \Gamma_{\rm eff}/dL$ of Eq.~(\ref{dGLNVeff}).
This oscillation term comes from the interference term in the square of 
amplitude (\ref{calAMpl}). Therefore, by comparing the obtained expression 
(\ref{dGLNVeffo1}) with (\ref{dGLNVeff}), we can obtain the complete
expression for the effective differential decay width with 
oscillation effects included
 \bea
\frac{d}{d L}
\Gamma_{\rm eff}^{\rm (osc)}(B^+ \to \mu^+ e^+ \pi^-;L) 
& \approx & 
\frac{1}{\gamma_N \beta_N}
\bG(B^{+} \to \mu^{+} N) \bG(N \to e^{+} \pi^{-})
\nonumber\\
&& \times
\left\{ \sum_{j=1}^2 |B_{\mu N_j}|^2 |B_{e N_j}|^2 +
2 {\rm Re} \left[ B_{\mu N_1}^* B_{e N_1}^* B_{\mu N_2} B_{e N_2}
\exp\left[  i (p_{N_2} - p_{N_1}) \cdot z \right] \right] \right\}.
\label{dGLNVeffo}
\eea
The oscillation term here contains two on-shell 4-momenta
$p_{N_j} = (E_{N_j},0,0,p_{N_j}^3)$ ($j=1,2$) which are related
by the on-shellness conditions $p_{N_j} \cdot p_{N_j} = M_{N_j}^2$ 
and by the condition
\be
\beta_{N_2} - \beta_{N_1} \equiv 
\frac{p_{N_2}^3}{E_{N_2}} - \frac{p_{N_1}^3}{E_{N_1}} \approx 0.
\label{delvj}
\ee
This condition comes from the following interpretation. The $N_1$ and $N_2$
amplitudes interfere at $L$ if both of them are appreciable there.
The neutrinos $N_1$ and $N_2$, in general, separate as they travel from their
production to their detection vertex. Interference is then possible
there only if this separation 
$|\Delta L_{12}| \equiv |(\beta_{N_2} - \beta_{N_1})| t$ 
(with: $t \approx L/\beta_N$) is smaller than 
the spread of the wave packet $\Delta L_{\rm wp} \equiv \beta_N \Delta T$, cf.~Ref.~\cite{Glash}
\be
\frac{|\beta_{N_2} - \beta_{N_1}|}{|\beta_{N_2} + \beta_{N_1}|} \ll 
\frac{\Delta T}{t} \; (\ll 1).
\label{lim}
\ee
\begin{figure}[htb]
\centering\includegraphics[width=120mm]{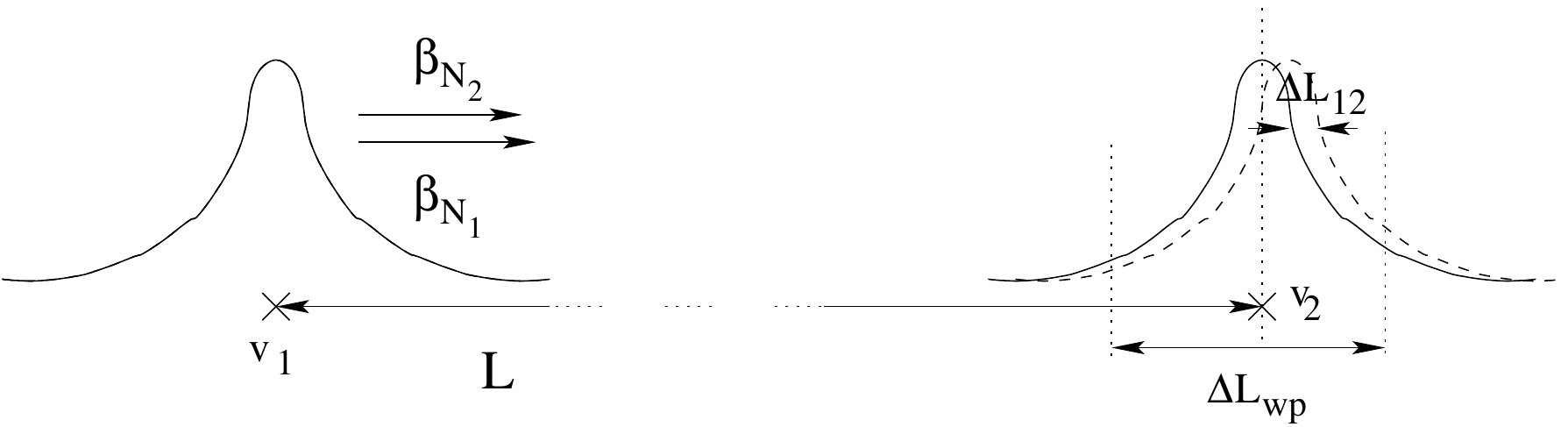}
\caption{Graphical representation of the hierarchy equation (\ref{hier})
of the lengths $\Delta L_{12}$, $\Delta L_{\rm wp}$ and
the detector length $L$. The two interaction vertices (the production and the
decay vertex of the neutrino $N$) are denoted as $v_1$ and $v_2$, respectively.
Note that at the production vertex ($v_1$) the wave packets of $N_1$ and $N_2$
are not mutually displaced, unlike in the decay vertex ($v_2$).}
\label{Lfig}
\end{figure}
Stated otherwise, the following hierarchy is assumed:
\be
|\Delta L_{12}| \left( \equiv \frac{|\beta_{N_2}-\beta_{N_1}| L}{\beta_N} \right)
\; \ll \; \Delta L_{\rm wp} \left(\equiv \beta_N \Delta T \right) \;  \ll \; L,
\label{hier}
\ee
cf.~also Fig.~\ref{Lfig}.

In order to express the oscillation phase 
$\phi(L)\equiv (p_{N_2} - p_{N_1}) \cdot z$ in
Eq.~(\ref{dGLNVeffo}) [$\Leftrightarrow$ (\ref{lim})] in a convenient form,
condition (\ref{delvj}) can be used. Since
$\beta_{N_2} \equiv p_{N_2}^3/E_{N_2}$ and $\beta_{N_1} \equiv p_{N_1}^3/E_{N_1}$
are close in value, hence they are close to the value of
$\beta_N \equiv (p_{N_2}^3+p_{N_1}^3)/(E_{N_2}+E_{N_1})$. It can be checked 
to see that
the latter velocity is practically equal to the arithmetic average 
$(1/2)(\beta_{N_2}+\beta_{N_1})$. Therefore,
\be
z = (t , 0,0,L)  \approx t (1, 0,0,\beta_N) = 
\frac{t}{(E_{N_2}+E_{N_1})} (p_{N_2}+p_{N_1})
\label{z}
\ee
Therefore, the oscillation phase is \cite{Glash}
\bea
\phi(L)\equiv (p_{N_2} - p_{N_1}) \cdot z &= &
t \frac{(M_{N_2}^2 - M_{N_1}^2)}{(E_{N_1}+E_{N_2})} \approx 
t M_N \frac{\Delta M_N}{E_N} 
\nonumber\\
&\approx& 
\frac{L}{\beta_N}  M_N \frac{\Delta M_N}{E_N} =
L \frac{\Delta M_N}{\beta_N \gamma_N},
\label{phiL}
\eea
where it was taken into account that $p_{N_j}^2=M_{N_j}^2$, and
$M_{N_2}^2 - M_{N_1}^2 = 2 M_N \Delta M_N$ 
(where $\Delta M_N \equiv M_{N_2} - M_{N_1}$ 
and $|\Delta M_N| \ll M_{N_1} \equiv M_{N}$). As stressed in 
Ref.~\cite{Glash}, this expression for the oscillation angle is valid always, 
not just for relativistic neutrinos $N_j$, whenever relation (\ref{lim})
is fulfilled. For example, if the neutrinos are nonrelativistic, 
we have $\phi(L) \approx (L/\beta_N) \Delta M_N$.
The obtained oscillation phase allows us to define the oscillation
length $L_{\rm osc}$ as 
\bea
\phi(L_{\rm osc}) & = & 2 \pi \; \Rightarrow \;
L_{\rm osc} = \frac{2 \pi \beta_N \gamma_N}{\Delta M_N}.
\label{Losc}
\eea
Using expression (\ref{phiL}), the differential decay width 
(\ref{dGLNVeffo}) can now be written in a more explicit form
 \bea
\lefteqn{
\frac{d}{d L}
\Gamma_{\rm eff}^{\rm (osc)}(B^+ \to \mu^+ e^+ \pi^-;L) 
 \approx 
\frac{1}{\gamma_N \beta_N}
\bG(B^{+} \to \mu^{+} N) \bG(N \to e^{+} \pi^{-})
}
\nonumber\\
&& \times
\left\{ \sum_{j=1}^2 |B_{\mu N_j}|^2 |B_{e N_j}|^2 +
2  |B_{\mu N_1}| |B_{e N_1}| |B_{\mu N_2}| |B_{e N_2}|
\cos\left( L \frac{\Delta M_N}{\beta_N \gamma_N} + \theta_{21}^{\rm (LNV)} \right)
\right\}
\label{dGLNVeffof}
\eea
where the constant phase $\theta_{21}^{\rm (LNV)}$ is defined in 
Eq.~(\ref{theta21LNV}).
We can integrate the differential decay width (\ref{dGLNVeffof})
over the $dL$ length to the full length $L$ between the vertices.
If $L \gg |L_{\rm osc}|$, this then gives
the full effective decay width  of Eq.~(\ref{GLNVeffb})
because the oscillation term $\sim \cos(\phi(L)+  \theta_{12})$
gives a relatively negligible contribution when integrated over several
``oscillation wavelengths'' $|L_{\rm osc}|$. If, on the other hand,
we do not assume $L \gg |L_{\rm osc}|$, the integration of
expression (\ref{dGLNVeffof}) gives
\bea
\Gamma_{\rm eff}^{\rm (osc)}(B^+ \to \mu^+ e^+ \pi^-;L) 
 & \approx &
\frac{L}{\gamma_N \beta_N}
\bG(B^{+} \to \mu^{+} N) \bG(N \to e^{+} \pi^{-})
{\bigg \{} \sum_{j=1}^2 |B_{\mu N_j}|^2 |B_{e N_j}|^2
\nonumber\\
&& 
+ \frac{L_{\rm osc}}{\pi L}  |B_{\mu N_1}| |B_{e N_1}| |B_{\mu N_2}| |B_{e N_2}|
\left[ \sin \left( 2 \pi \frac{L}{L_{\rm osc}} + \theta_{21}^{\rm (LNV)} \right)
- \sin ( \theta_{21}^{\rm (LNV)} ) \right]
{\bigg \}}.
\label{GLNVeffof}
\eea

Until now we considered the case of oscillation effects in LNV decays
$B^+ \to \mu^+ e^+ \pi^-$. It can be checked that for the charge-conjugate
LNV decays $B^- \to \mu^- e^- \pi^+$ the previous derivation can be repeated,
with the only replacements $B_{\ell N_j}^* \mapsto B_{\ell N_j}$ and
$B_{\ell N_j} \mapsto B_{\ell N_j}^*$. Instead of Eq.~(\ref{calAMpl})
we now have
\be
{\cal A}(B^- \to \mu^- e^- \pi^{+}; L) \sim 
B_{\mu N_1} B_{e N_1} \exp(- i p_{N_1} \cdot z) +
B_{\mu N_2} B_{e N_2} \exp(- i p_{N_2} \cdot z).
\label{calAMmi}
\ee 
This implies that in result (\ref{dGLNVeffof}) we now get
$\theta_{21} \mapsto -\theta_{21}$, so that we can extend the
results (\ref{dGLNVeffof}) and (\ref{GLNVeffof})
to both LNV cases ($B^{\pm}$)
 \bea
\lefteqn{
\frac{d}{d L}
\Gamma_{\rm eff}^{\rm (osc)}(B^{\pm} \to \mu^{\pm} e^{\pm} \pi^{\mp};L) 
 \approx  
\frac{1}{\gamma_N \beta_N}
\bG(B^{+} \to \mu^{+} N) \bG(N \to e^{+} \pi^{-})
}
\nonumber\\
&& \times
\left\{ \sum_{j=1}^2 |B_{\mu N_j}|^2 |B_{e N_j}|^2 +
2  |B_{\mu N_1}| |B_{e N_1}| |B_{\mu N_2}| |B_{e N_2}|
\cos\left(  2 \pi \frac{L}{L_{\rm osc}} \pm \theta_{21}^{\rm (LNV)} \right)
\right\},
\label{dGLNVeffofpm}
\eea
\bea
\Gamma_{\rm eff}^{\rm (osc)}(B^{\pm} \to \mu^{\pm} e^{\pm} \pi^{\mp};L) 
 & \approx &
\frac{L}{\gamma_N \beta_N}
\bG(B^{+} \to \mu^{+} N) \bG(N \to e^{+} \pi^{-})
{\bigg \{} \sum_{j=1}^2 |B_{\mu N_j}|^2 |B_{e N_j}|^2 
\nonumber\\
&& 
+\frac{L_{\rm osc}}{\pi L}  |B_{\mu N_1}| |B_{e N_1}| |B_{\mu N_2}| |B_{e N_2}|
\left[ \sin \left( 2 \pi \frac{L}{L_{\rm osc}} \pm \theta_{21}^{\rm (LNV)} \right)
\mp \sin ( \theta_{21}^{\rm (LNV)} ) \right]
{\bigg \}}.
\label{GLNVeffofpm}
\eea

For the LNC processes $B^{\pm} \to \mu^{\pm} e^{\mp} \pi^{\pm}$
(cf.~Fig.~\ref{FigBmuepiLNC}), in the case of no oscillation effects the
results for the decay widths are the same as for the LNV processes; 
cf.~Eqs.~(\ref{GLNVeff})-(\ref{LNCeff}). When oscillations are accounted for,
the results are almost the same as in the just considered LNV processes,
except that for the decay amplitudes 
[cf.~Eqs.~(\ref{calAMpl}) and (\ref{calAMmi}) for LNV case]
we have some of the heavy-light mixing elements $B_{\ell N_j}$
complex-conjugated and others not
\bes
\label{calAMLNC}
\bea
{\cal A}(B^+ \to \mu^+ e^- \pi^{+}; L) &\sim& 
B_{\mu N_1}^* B_{e N_1} \exp(- i p_{N_1} \cdot z) +
B_{\mu N_2}^* B_{e N_2} \exp(- i p_{N_2} \cdot z),
\label{calAMLNCpl}
\\
{\cal A}(B^- \to \mu^- e^+ \pi^{-}; L) &\sim& 
B_{\mu N_1} B_{e N_1}^* \exp(- i p_{N_1} \cdot z) +
B_{\mu N_2} B_{e N_2}^* \exp(- i p_{N_2} \cdot z).
\label{calAMLNCmi}
\eea
\ees
This then leads to the following results, in analogy with the
LNC results (\ref{dGLNVeffofpm})-(\ref{GLNVeffofpm}) where now only
the phase angle $\theta_{21}^{\rm (LNV)}$ gets replaced by a different phase
angle $\theta_{21}^{\rm (LNC)}$ given in Eq.~(\ref{theta21LNC}):
 \bea
\frac{d}{d L}
\Gamma_{\rm eff}^{\rm (osc)}(B^{\pm} \to \mu^{\pm} e^{\mp} \pi^{\pm};L) 
 &\approx &
\frac{1}{\gamma_N \beta_N}
\bG(B^{+} \to \mu^{+} N) \bG(N \to e^{+} \pi^{-})
\nonumber\\
&& \times
\left\{ \sum_{j=1}^2 |B_{\mu N_j}|^2 |B_{e N_j}|^2 +
2  |B_{\mu N_1}| |B_{e N_1}| |B_{\mu N_2}| |B_{e N_2}|
\cos\left( 2 \pi \frac{L}{L_{\rm osc}} \pm \theta_{21}^{\rm (LNC)} \right)
\right\},
\label{dGLNCeffofpm}
\eea
\bea
\Gamma_{\rm eff}^{\rm (osc)}(B^{\pm} \to \mu^{\pm} e^{\mp} \pi^{\mp};L) 
 & \approx &
\frac{L}{\gamma_N \beta_N}
\bG(B^{+} \to \mu^{+} N) \bG(N \to e^{+} \pi^{-})
{\bigg \{} \sum_{j=1}^2 |B_{\mu N_j}|^2 |B_{e N_j}|^2 
\nonumber\\
&& 
+\frac{L_{\rm osc}}{\pi L}  |B_{\mu N_1}| |B_{e N_1}| |B_{\mu N_2}| |B_{e N_2}|
\left[ \sin \left( 2 \pi \frac{L}{L_{\rm osc}} \pm \theta_{21}^{\rm (LNC)} \right)
\mp \sin ( \theta_{21}^{\rm (LNC)} ) \right]
{\bigg \}}.
\label{GLNCeffofpm}
\eea

All the formulas with oscillation effects, derived in this section,
can be extended in a straightforward way
to the oscillation effects in the semihadronic decays with two equal flavors
of produced charged leptons, i.e., $M^{\pm} \to \ell^{\pm} \ell^{\pm} M^{' \mp}$
and $M^{\pm} \to \ell^{\pm} \ell^{\mp} M^{' \pm}$;
more specifically,  $B^{\pm} \to \mu^{\pm} \mu^{\pm} \pi^{\mp}$
and  $B^{\pm} \to \mu^{\pm} \mu^{\mp} \pi^{\pm}$ (cf.~Sec.~\ref{sec:B}).

In Appendix we show that the wave function approach
of Ref.~\cite{Bilenky} (cf.~also \cite{Giunti}) to the considered
LNV and LNC processes with on-shell neutrinos is consistent, within
their approximations, with the amplitude approach presented here and
based on the method of Ref.~\cite{Glash}. 

A question may appear why the usual ($S$-matrix) approach, leading to
the results of the Sec.~\ref{sec:gamma}, did not give the
modulation (oscillation) effects obtained in this section. 
The $x$-coordinates of fields are integrated over in the 
$S$-matrix approach of Sec.~\ref{sec:gamma}, reflected
by the use of initial and final states with
definite momenta ($\delta p = 0$). The uncertainty relation implies then
$\delta x = \infty$. Therefore, the location of the
vertices remained undefined in the approach of the previous section.
On the other hand, if the location of the vertices is to 
be determined in an experiment by precision $\delta x (= \delta L) \sim
0.1$ mm or better, then the corresponding precision in the determination
of the momenta is $\delta p \agt 1/\delta x \sim 10^{-4}$ eV.

\section{Oscillation length and measurement of the modulation}
\label{sec:est}

For the described oscillation modulation to be well defined and detectable, 
several conditions have to be fulfilled. 
Among them is the hierarchy (\ref{hier}) between the length $L$ 
(the distance between the production and the decay vertices),
the width $L_{\rm wp}$ of the wave packet, and the separation $\Delta L_{12}$
between the two wave packets at the second vertex (cf.~Fig.~\ref{Lfig}
and Ref.~\cite{Glash}). Yet another necessary condition for the
detection of the oscillation is that the maximal detected length $L$ 
between the two vertices (we will call it simply the total detector 
length, $L_{\rm max} \equiv L_{\rm det}$) is larger than or comparable with the
oscillation length $|L_{\rm osc}|$ [Eq.~(\ref{Losc})]. For the measurement
of the oscillation modulation effects in practice, the case
$|L_{\rm osc}| \sim L_{\rm det}$ is more convenient
than $|L_{\rm osc}| < L_{\rm det}$, i.e.,
\be
|L_{\rm osc}| \ \left( \equiv \frac{2 \pi \beta_N \gamma_N}{|\Delta M_N|} \right)
\; \sim \; L_{\rm max} \ (\equiv L_{\rm det}).
\label{LoscL}
\ee
Furthermore, if the decay probability $P_{N_j}(L_{\rm det})$ for the decay of
$N_j$ ($j=1,2$) within the detector [Eqs.~(\ref{PN}) and (\ref{bPNj})]
is significant, i.e., if  $P_{N_j}(L_{\rm det}) \sim 1$, then the oscillation
is not well defined because it disappears within one or less oscillation cycle
due to the decay of $N_j$. Therefore, for the oscillation to be 
well defined, we have to require $P_{N_j}(L_{\rm det}) \ll 1$. This
means, according to Eq.~(\ref{PNb}) and using Eq.~(\ref{LoscL}),
the following:
\be
\left( |L_{\rm osc}| \equiv \right) 
\frac{2 \pi \beta_N \gamma_N}{|\Delta M_N|} \sim L_{\rm det} \ll
\frac{\beta_N \gamma_N}{\Gamma_{N_j}}.
\label{hier2}
\ee
This implies that we have $1/|\Delta M_N| \ll 1/\Gamma_{N_j}$ ($j=1,2$), 
meaning that the condition $|y| (\equiv |\Delta M_N|/\Gamma_N) \gg 1$ of
Eq.~(\ref{cond}) is fulfilled when we have well-defined and detectable
oscillation.\footnote{$\Gamma_N \equiv (1/2)(\Gamma_{N_1}+\Gamma_{N_2})$
according to definition (\ref{GNy}).}
We recall that this condition ($|y| \gg 1$) was
assumed throughout the derivation of the oscillation formulas of the previous 
section so that the (otherwise problematic) overlap terms with 
$\delta(y)$ and $\eta(y)/y$ factors in expression (\ref{GDDLNV})
could be neglected.

The oscillation length can be estimated in the following way. 
Let us assume that the near mass  degeneracy ($y \gg 1$) is in the
interval: $1 \ll |y| (\equiv |\Delta M_N|/\Gamma_N) \lesssim 10^2$,
i.e.,\footnote{We do not want to have $|y| > 10^2$ because the
CP violation effects are then very suppressed \cite{CKZ2}.}
\be
|\Delta M_N| \lesssim 10^2 \Gamma_N.
\label{dMNub}
\ee
Furthermore, let us take that in the total decay widths $\Gamma_{N_j}$,
Eqs.~(\ref{GNwidth})-(\ref{calK}), the dominating contribution
in the mixing factors $\K_j$ is from $\ell$-component, i.e.,
$\K_j \approx {\cal N}_{\ell N} |B_{\ell N_j}|^2$ ($j=1,2$; $\ell=e$ or $\mu$ or
$\tau$). Stated otherwise, we assume that $|B_{\ell N_j}|^2$ is the largest
among the mixings $|B_{e N_j}|^2$, $|B_{\mu N_j}|^2$ and $|B_{\tau N_j}|^2$. 
Then, we have
\bea
\Gamma_{N_j} &=& \left( \frac{{\cal N}_{\ell N}}{10} \right) \times
\left( \frac{ |B_{\ell N_j}|^2 }{ 10^{-5}} \right) \times 4.57 \times 10^{-18} \ {\rm GeV}
\nonumber\\
& = & \left( \frac{{\cal N}_{\ell N}}{10} \right) \times
\left( \frac{ |B_{\ell N_j}|^2 }{ 10^{-5}} \right) \times \frac{1}{43.5 \ {\rm m}}.
\label{GNjinm}
\eea
For $M_N =1$-$5$ GeV, and taking $N$ to be Majorana neutrino,
we have ${\cal N}_{e N} \approx {\cal N}_{\mu N} 
\approx 6$-$10$, and ${\cal N}_{\tau N} \approx 3$-$5$ 
(cf.~Refs.~\cite{CKZ2,CDKZ}); hence, the factor ${\cal N}_{\ell N}/10$ in
Eq.~(\ref{GNjinm}) is $\sim 1$. The factor $|B_{\ell N_j}|^2/10^{-5}$
in Eq.~(\ref{GNjinm}) can be $\sim 1$, or larger or smaller;
cf.~Table \ref{B2ub} for some present upper bounds.
\begin{table}
\centering
\caption{Presently known upper bounds for the squares $|B_{\ell N}|^2$ of the
heavy-light mixing matrix elements, for various specific values of $M_N$.
We excluded the upper bounds for $B_{e N}|^2$ from the $0\nu\beta\beta$ decay, 
which are uncertain due to possible cancellation effects. See also 
Figs.~$3$-$5$ of Ref.~\cite{DBP}. For each upper bound, the corresponding 
experiment (reference) is indicated.}
\label{B2ub}
\begin{tabular}{| c | c | c | c |}
\hline
\bf{$M_N [GeV]$} & $|B_{eN}|^2$ & $|B_{\mu N}|^2$ & $|B_{\tau N}|^2$ \\
\hline
1.0 & $3\times10^{-7}$ (\cite{charm}) & $ 1 \times10^{-7}$ (\cite{NuTeV}) & $3\times10^{-3}$ (\cite{delphi}) \\
\hline
2.1 & $4 \times10^{-5}$ (\cite{delphi})& $3\times10^{-5}$ (\cite{Belle})& $2\times10^{-4}$ (\cite{delphi}) \\
\hline
3.0 & $2\times10^{-5}$(\cite{delphi}) & $2\times10^{-5}$(\cite{delphi}) & $4\times10^{-5}$(\cite{delphi}) \\
\hline
$4.0$-$5.0$ & $1\times10^{-5}$ (\cite{delphi})& $1\times10^{-5}$ (\cite{delphi}) & $1\times10^{-5}$ (\cite{delphi} \\
\hline
\end{tabular}
\end{table}
The oscillation length (\ref{Losc}) can then be estimated
\bes
\label{Losces}
\bea
|L_{\rm osc}| &=& \frac{2 \pi |{\vec p}_N|}{M_N |\Delta M_N|} \gtrsim 
\frac{ |{\vec p}_N|}{M_N} \frac{2 \pi}{10^2 \Gamma_N} \sim
 \frac{ |{\vec p}_N|}{M_N} \frac{1}{10 \Gamma_N}
\label{Loscesa}
\\
& \sim &  \frac{ |{\vec p}_N|}{M_N} \frac{1}{10} 
\times \left( \frac{10}{{\cal N}_{\ell N}} \right) \times
\frac{2 \times 10^{-5}}{(|B_{\ell N_1}|^2 + |B_{\ell N_2}|^2)} \times 5 \times 10^1 \ {\rm m} 
\sim  \frac{ |{\vec p}_N|}{M_N} \frac{10^{-4}}{|B_{\ell N_j}|^2} \ {\rm m}
\sim  \frac{10^{-4}}{|B_{\ell N_j}|^2} \ {\rm m}.
\label{Loscesb}
\eea
\ees
In estimate (\ref{Loscesa}) we assumed inequality (\ref{dMNub}),
and in estimate  (\ref{Loscesb}) we took into account
relation (\ref{GNjinm}), as well as identity (\ref{GNy}) for $\Gamma_N$;
at the end, we assumed that the produced on-shell neutrinos $N_j$ are
semirelativistic, i.e., $|{\vec p}_N| \sim M_N$ ($\sim 1$ GeV).
Using estimate (\ref{Loscesb}) and recalling that
$|B_{\ell N_j}|^2$ is the largest
among the mixings $|B_{e N_j}|^2$, $|B_{\mu N_j}|^2$ and $|B_{\tau N_j}|^2$,
we can see from Table \ref{B2ub} 
that for $M_N = 1$-$5$ GeV we can take $|B_{\ell N_j}|^2 = 
|B_{\tau N_j}|^2$, whose upper bounds are given in the right column
of Table \ref{B2ub}. This implies that, at present, we can expect
the values $L_{\rm osc} \sim 0.1$-$10$ m for the oscillation length.
Of course, implicitly we assumed that the energies of the
($B$ or $B_c$) mesons, which decay, are not very high so that the assumption
$|{\vec p}_N| \sim M_N$ would be justified. If $|L_{\rm osc}| > 10$ m, we would
need quite a large detector, cf.~Eqs.~(\ref{LoscL}) and (\ref{hier2}).

If we have $|L_{\rm osc}| \sim 0.1$-$1$ m ($\sim L_{\rm det}$), our formulas 
(\ref{dGLNVeffofpm})-(\ref{GLNVeffofpm}) for LNV decays and 
(\ref{dGLNCeffofpm})-(\ref{GLNCeffofpm}) for LNC decays indicate
that such oscillations can be detected and measured, once a sufficient
number of such decays is detected, with the first (production) and
the second (decay) vertices being within the detector. In this
way, the oscillation length $L_{\rm osc} \propto 1/\Delta M_N$ could
be determined, and thus the mass difference $\Delta M_N$ ($\ll M_N$). 

It is also interesting that these formulas indicate that in such a case
the phases $\theta_{21}^{\rm (LNV)}$ and $\theta_{21}^{\rm (LNC)}$ could be measured
as well. These phases could be determined, for example, by comparing
the modulation of the measured differential effective decay widths
$d \Gamma_{\rm eff}^{\rm (osc)}(B^{\pm};L)/dL$ for the $B^+$ and $B^-$
decays into $\mu e \pi$, because the phase difference between the
two oscillatory modulations is $2 \times \theta_{21}$; 
cf.~Eq.~(\ref{dGLNVeffofpm}) for the LNV and Eq.~(\ref{dGLNCeffofpm})
for the LNC case.
The factor $\sin \theta_{21}$ appears in the CP asymmetry factor 
${\cal A}_{\rm CP} \propto \sin \theta_{21}$ for these processes.
For example, this asymmetry for the LNV case is
\bes
\label{ACP}
\bea
{\cal A}_{\rm CP}^{\rm (LNV)}(B)  &\equiv&
\frac{\Gamma(B^- \to \mu^- e^- \pi^{+}) -\Gamma(B^+ \to \mu^+ e^+ \pi^{ -})}
{\Gamma(B^- \to \mu^- e^- \pi^{+}) + \Gamma(B^+ \to \mu^+ e^+ \pi^{-})}
\label{ACPdef}
\\
& \propto & P \sin \theta_{21}^{\rm (LNV)} \frac{y}{y^2 + 1},
\eea
\ees
where $y \equiv \Delta M_N/\Gamma_N$ [cf.~notation (\ref{GNy})], and
factor $P \sim 1$ depends principally on the ratios of
mixings $|B_{\ell N_2}|/|B_{\ell N_1}|$ ($\ell = \mu, e$) and ratio
$\K_1/\K_2$ [cf.~notation (\ref{calK})].
This factor ${\cal A}_{\rm CP}$ can be substantial if 
$y \equiv \Delta M_N/\Gamma_N$ is not too small in absolute value, 
e.g. if $|y| \sim 10$.
We refer to Refs.~\cite{CKZ2,CDKZ} for more details on this.
An interesting aspect here is that, by the described measurement of
the angle $\theta_{21}$ we could conclude that the CP asymmetry
${\cal A}_{\rm CP}$ is nonzero even in the case when $|y| \gg 1$, i.e., 
when this asymmetry is practically unmeasurable.

\begin{figure}[htb]
\centering\includegraphics[width=100mm]{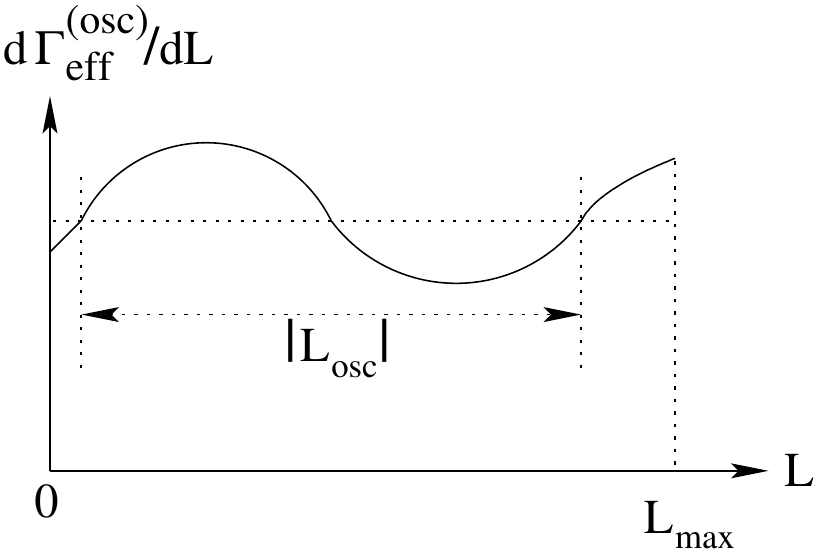}
\caption{Graphical schematical representation of the differential decay rate
$d \Gamma_{\rm eff}^{\rm (osc)}(B \to \mu e \pi; L)/dL$, cf.~Eqs.~(\ref{dGLNVeffofpm}) and (\ref{dGLNCeffofpm}).}
\label{dGdLfig}
\end{figure}
The differential decay width $d \Gamma_{\rm eff}^{\rm (osc)}(B \to \mu e \pi; L)/dL$
of Eqs.~(\ref{dGLNVeffofpm}) and (\ref{dGLNCeffofpm}) is presented 
schematically in Fig.~\ref{dGdLfig}, where $L$ is the distance between the 
two vertices and $L_{\rm max} = L_{\rm det}$. In order to interpret how to measure
this differential decay width, we recall that this quantity is the limit 
$(1/\Delta L) \times \Gamma_{\rm eff}^{\rm (osc)}(B \to \mu e \pi; L < L^{'} < L + \Delta L)$ when $\Delta L \to +0$ (i.e., $\Delta L \ll |L_{\rm osc}|$), and here
$L^{'}$ is the distance between the production ($\mu$-$N_j$)
and the decay ($N_j$-$e$-$\pi$) vertex.
To measure such a quantity, a sufficiently high number of events for each
chosen bin $L < L' < L + \Delta L$ would have to be measured
(with $\Delta L \ll |L_{\rm osc}|$ and $L \leq L_{\rm det}$).

There may exist another complication in such measurements. Namely the length
$L_{\rm osc}$ can vary in the detected events of the considered decays
because $L_{\rm osc} \propto \beta_N \gamma_N \propto |{\vec p}_N| \equiv 
|{\vec p}_e + {\vec p}_{\pi}|$.  In principle, the 3-momentum ${\vec p}_N 
\equiv {\vec p}_e + {\vec p}_{\pi}$ can be measured in each such decay,
i.e., $L_{\rm osc}$ can be determined in each such event. The graphical 
representation Fig.~\ref{dGdLfig} refers to a class of events which,
among themselves, have approximately equal value of $L_{\rm osc}$, i.e.,
approximately equal $|{\vec p}_N|$. If the decaying $B^{\pm}$ (or
$B_c^{\pm}$) mesons were at rest in the lab frame, then the value of
$|{\vec p}_N|$ is such a frame would be fixed by kinematics, namely
\be
|{\vec p}_N^{(0)}| = \frac{1}{2} M_B \; \lambda^{1/2}\left(1,
\frac{M_N^2}{M_B^2}, \frac{M_{\mu}^2}{M_B^2} \right),
\label{pNrest}
\ee
where notation (\ref{lambdaN}) is used. 

In reality the $B$ mesons coming into the detector have energies 
$E_B > M_B$. Let us assume that the incoming $B$ mesons in the lab frame 
have all approximately the same 3-momentum ${\vec p}_B = |{\vec p}_B| {\hat z}$ 
parallel to the direction ${\hat z}$ of the tube of 
the detector where both vertices are detected, and that the detector tube 
is relatively narrow. Then the vector ${\vec p}_N$ in the detected events
is the 3-momentum which can be obtained from 
${\vec p}_N^{(0)} = |{\vec p}_N^{(0)}| {\hat z}$ [cf.~Eq.~(\ref{pNrest})]
by a constant boost in the direction $-{\hat z}$, bringing us 
from the $B$ rest frame into the lab frame where $B$'s have the 
(approximately) constant 3-momentum $|{\vec p}_B| {\hat z}$. 
Thus the lab 3-momentum ${\vec p}_N = |{\vec p}_N| {\hat z}$ is
approximately constant also in such a case. In such a case $L_{\rm osc}$
would be approximately the same for all the detected events $B \to \mu N
\to \mu e \pi$ in the tube, and the oscillation modulation 
indicated in Fig.~\ref{dGdLfig} could be measured, including the
phase $\theta_{21}$ relevant to CP violation.

\section{Determination of the heavy-light mixing coefficients $|B_{\ell N_j}|^2$.}
\label{sec:B}

Measurement of the differential effective decay widths
$d \Gamma_{\rm eff}^{\rm (osc)}(B^{\pm};L)/dL$ can also lead to
determination of the absolute values $|B_{\ell N_1}|$ and
$|B_{\ell N_2}|$ of the heavy-light mixing coefficients.
For example, for determination of $|B_{\mu N_j}|$ ($j=1,2$) 
it is convenient to consider the decay widths for the 
semileptonic LNV decays $B^{\pm} \to \mu^{\pm} \mu^{\pm} \pi^{\mp}$
(and/or for the LNC variant)
\bea
\frac{d}{d L}
\Gamma_{\rm eff}^{\rm (osc)}(B^{\pm} \to \mu^{\pm} \mu^{\pm} \pi^{\mp};L) 
 & \approx & 
\frac{1}{\gamma_N \beta_N}
\bG(B^{+} \to \mu^{+} N) \bG(N \to \mu^{+} \pi^{-})
\nonumber\\
&& \times
\left\{ |B_{\mu N_1}|^4 + |B_{\mu N_2}|^4 +
2  |B_{\mu N_1}|^2 |B_{\mu N_2}|^2
\cos\left(  2 \pi \frac{L}{L_{\rm osc}} \pm \phi_{21}^{\rm (LNV)} \right)
\right\},
\label{dGLNVeffmumu}
\eea
where $\phi_{21}^{\rm (LNV)} = 2 {\rm arg}(B_{\mu N_2}) - 2 {\rm arg}(B_{\mu N_1})$,
in analogy with the $\mu e$ case Eq.~(\ref{theta21LNV}).
Expression (\ref{dGLNVeffmumu}) differs from expression 
(\ref{dGLNVeffofpm}) only by the replacements $e \mapsto \mu$
(and $\theta_{21} \mapsto \phi_{21}$).
In the case of $B^{\pm} \to \mu^{\pm} \mu^{\pm} \pi^{\mp}$, 
the symmetry factor $(1/2!)$ due to two identical muons in the final state
gets canceled by the factor $2$ coming from the inclusion of the
square of the crossed channel amplitude, cf.~Ref.~\cite{CKZ2}. The crossed
channel amplitude in the decay $B^{\pm} \to \mu^{\pm} e^{\pm} \pi^{\mp}$
did not enter because it represents a different (distinguishable)
process once the two vertices are identified and localized in the experiment.
Stated otherwise, the processes 
$B^{\pm} \to \mu^{\pm} N \to \mu^{\pm} e^{\pm} \pi^{\mp}$ and
$B^{\pm} \to e^{\pm} N \to e^{\pm} \mu^{\pm} \pi^{\mp}$
are distinguishable once the vertices are identified.

Measurement of the average (over $L$) 
$\langle d \Gamma_{\rm eff}^{\rm (osc)}/d L \rangle$ 
and of the modulation amplitude $\Delta ( d \Gamma_{\rm eff}^{\rm (osc)}/d L )$
determines the quantities (see also Fig.~\ref{dGdLfig})
\be
|B_{\mu N_1}|^4 + |B_{\mu N_2}|^4 \equiv \langle F \rangle,
\quad
2  |B_{\mu N_1}|^2 |B_{\mu N_2}|^2 \equiv \Delta F,
\label{FDF}
\ee
respectively. Let us denote as $N_1$ the neutrino with larger mixing element
($|B_{\mu N_1}| > |B_{\mu N_2}|$).\footnote{
We have the freedom to do that, 
because the formulas are symmetric under the exchange
$N_1 \leftrightarrow N_2$ (i.e., $\phi_{21} \mapsto - \phi_{21}$; 
$\Delta M_N \mapsto - \Delta M_N$, $L_{\rm osc} \mapsto - L_{\rm osc}$).}
Then the heavy-light mixing coefficients $|B_{\mu N_j}|^2$ ($j=1,2$)
are determined as well
\bes
\label{BmuN2}
\bea
|B_{\mu N_1}|^2 & = & \frac{1}{2} \left(
\sqrt{\langle F \rangle + \Delta F} + \sqrt{\langle F \rangle - \Delta F}
\right),
\\
|B_{\mu N_2}|^2 & = & \frac{1}{2} \left(
\sqrt{\langle F \rangle + \Delta F} - \sqrt{\langle F \rangle - \Delta F}
\right),
\eea
\ees

\section{Conclusions}
\label{sec:concl}

In this work we considered the phenomenon of neutrino oscillations
in semileptonic decays of $B$ mesons
via on-shell heavy nearly mass-degenerate Majorana neutrinos $N_j$ ($j=1,2$):
the lepton number violating (LNV) decays
$B^{\pm} \to \mu^{\pm} N_j \to \mu^{\pm} e^{\pm} \pi^{\mp}$, 
and the lepton number conserving (LNC) decays
$B^{\pm} \to \mu^{\pm} N_j \to \mu^{\pm} e^{\mp} \pi^{\pm}$.
Since the neutrinos contributing to such decays have to be on shell
(the off-shell neutrinos give completely negligible contributions),
the relevant flavor analogs are not $\nu_{\mu}$ and $\nu_e$ [Eq.~(\ref{mix})],
but the truncated combinations ${\cal N}_1$ and ${\cal N}_2$ 
[Eqs.~(\ref{ellWN}) and (\ref{cNs})], which are combinations of only
the heavy mass neutrinos $N_1$ and $N_2$. The central results of
the work are Eqs.~(\ref{dGLNVeffofpm}) and (\ref{dGLNCeffofpm})
for the LNV and LNC differential effective decay rates 
$d \Gamma_{\rm eff}^{\rm (osc)}(L)/d L$. These quantities must be interpreted as 
$(1/\Delta L) \times \Gamma_{\rm eff}^{\rm (osc)}(L < L' < L + \Delta L)$,
where $L'$ is the measured distance between the production vertex
($\mu$-$N$) and the decay vertex ($N$-$e$-$\pi$), and $\Delta L$
is considerably smaller than the oscillation length $|L_{\rm osc}| \equiv
2 \pi |{\vec p}_N|/( M_N |\Delta M_N|)$. Here, ${\vec p}_N$ is the 
(approximately constant) 3-momentum of the intermediate $N_j$'s, 
and mass quantities are $\Delta M_N \equiv M_{N_2} - M_{N_1}$ 
where $|\Delta M_N| \ll M_{N_1} \equiv M_N$.
We argued that it is possible to have $|L_{\rm osc}| \sim 0.1$-$10$ m
if the 3-momenta ${\vec p}_B$ and thus ${\vec p}_N$ are not too large.
If the detector length is comparable with $|L_{\rm osc}|$, and
a sufficient number of mentioned decays is detected,
we argued that it will be conceivable to measure the
$L$-dependence of the differential decay width
$d \Gamma_{\rm eff}^{\rm (osc)}(L)/d L$, i.e., the oscillation
modulation effects. By measuring these effects, the value of
$L_{\rm osc}$ could be discerned and thus the value of the mass
difference $\Delta M_N$.
Moreover, by measuring such effects
it would be possible to discern the phase $\theta_{21}$
[cf.~Eq.~(\ref{theta21LNV}) and (\ref{theta21LNC})], which plays
an important role in the CP violation.
In addition, magnitudes $|B_{\ell N_1}|$ and $|B_{\ell N_2}|$
($\ell = \mu, e$) of the heavy-light mixing coefficients
could be measured.

In all the presented formulas, we can replace the initial meson $B^{\pm}$
by any other heavy pseudoscalar meson (such as $B_c^{\pm}, D_s^{\pm}, B^{\pm}$),
and final meson $\pi^{\pm}$ by any other lighter meson (such as
$K^{\pm}, D^{\pm}, D_s^{\pm}$). This is performed by simply replacing 
everywhere the meson masses, the decay constants and
the CKM elements accordingly [cf.~Eqs.(\ref{notGDDN}), etc.].
Among the initial mesons, those which can be copiously produced
are evidently preferred. Additionally, those with higher mass are 
preferred because then the masses $M_N$ of the on-shell neutrinos 
can be larger; 
thus, the probability for the decay within the detector, $P_N(L)$,
can be more significant. Also, the preferred initial mesons are those
which have less CKM suppression, i.e., whose CKM element
$|V_{Q_u Q_d}|$ in Eq.~(\ref{K2}) is not too small. Therefore,
the preferred initial mesons $M^{\pm}$ are, in general, 
$D_s^{\pm}$ ($|V_{cs}| \approx 1$ and $M_{D_s} = 1.97$ GeV) 
and $B_c^{\pm}$ ($|V_{cb}| \approx 0.04$ and $M_{B_c} = 6.28$ GeV),
but not necessarily $B^{\pm}$ ($|V_{ub}| \approx 0.004$ and $M_B=5.28$ GeV).

Furthermore, as shown in Sec.~\ref{sec:B},
the obtained formulas can be extended in a straightforward
way to the decays in which the two charged leptons are equal,
i.e., $M^{\pm} \to \mu^{\pm} N_j \to \mu^{\pm} \mu^{\pm} \pi^{\mp}$ 
and   $M^{\pm} \to \mu^{\pm} N_j \to \mu^{\pm} \mu^{\mp} \pi^{\pm}$.

\begin{acknowledgments}
\noindent
This work was supported in part by FONDECYT (Chile) Grant No.~1130599 (G.C.), 
and Project Mecesup (Chile) FSM 1204 (J.Z.S.).
The work of C.S.K. was supported in part by the NRF grant funded by the Korean Government of the MEST (Grants No. 2011-0017430 and No. 2011-0020333).
\end{acknowledgments}

\appendix

\section{The quantum mechanics approach to oscillation}
\label{app:wf}

In this appendix, we show that the amplitude approach to 
on-shell oscillations in the considered processes, 
as presented in the main text of this work
and following mainly the amplitude approach of Ref.~\cite{Glash},
is consistent with the usual (quantum mechanics) approach
to neutrino oscillation \cite{Bilenky} (cf.~also \cite{Giunti})
applied to these processes (within the approximations
used in such approaches).

We recall that the relevant $e$- and $\mu$-flavor analogs in the
considered processes are the combinations (\ref{cN1N2})
of only the two almost mass-degenerate heavy neutrino eigenfields $N_j$ 
($j=1,2$), because
the other components (including the light neutrino mass eigenfields
$\nu_1$, $\nu_2$, $\nu_3$) are off shell or are assumed to be off shell
in the considered processes. 
Following the usual (quantum mechanics) approaches to neutrino oscillation,
cf.~\cite{Bilenky} (cf.~also \cite{Giunti}), the $e$ and $\mu$ heavy 
flavor analogs
${\cal N}_{\alpha}$ ($\alpha=1,2$) of the heavy neutrino mass eigenstates
$N_j$ ($j=1,2$) [cf.~Eq.~(\ref{ellWN})] are represented as quantum mechanical 
states [cf.~Eq.~(\ref{cN1N2}) for the corresponding fields]
\bes
\label{wfs}
\bea
| {\cal N}_{\alpha} \rangle & = & \B_{\alpha 1}^* | N_1 \rangle + \B_{\alpha 2}^* | N_2 \rangle,
\label{wfsa}
\\
| {\overline {\cal N}_{\alpha}} \rangle & = & \B_{\alpha 1} | N_1 \rangle + \B_{\alpha 2} | N_2 \rangle \quad (\alpha=1,2),
\label{wfsb}
\eea
\ees
where in Eq.~(\ref{wfsb}) we assume that the physical
neutrinos $N_j$ are Majorana. Here, we use notation (\ref{bB}) for
the $2 \times 2$ matrix $\B$ with normalized lines.
In the wave function approach \cite{Bilenky}, these wave functions are
in the Schr\"odinger representation and, consequently, have the following 
evolution in time $t$:
\bes
\label{tNt}
\bea
| {\cal N}_{\alpha}(t) \rangle & = & 
\sum_{j=1}^2 \B_{\alpha j}^* \exp(-i E_j t) | N_j \rangle
=  \sum_{\beta=1}^2 \sum_{j=1}^2 \B_{\alpha j}^*  \exp(-i E_j t)
\left( \B^{* -1} \right)_{j \beta} | {\cal N}_{\beta} \rangle,
\label{tNta}
\\
| {\overline {\cal N}_{\alpha}}(t) \rangle & = & 
\sum_{j=1}^2 \B_{\alpha j} \exp(-i E_j t) | N_j \rangle
=  \sum_{\beta=1}^2 \sum_{j=1}^2 \B_{\alpha j}  \exp(-i E_j t)
\left( \B^{ -1} \right)_{j \beta} | {\overline {\cal N}_{\beta}} \rangle,
\label{tNtb}
\eea
\ees
where we recall notation (\ref{bB}) used for the $2 \times 2$ matrix $\B$,
and the inverse matrix is, consequently,
\be
\B^{-1} = \frac{1}{{\rm Det} \B}
\left[
\begin{array}{rr}
\B_{22} & - \B_{12}
\\
-\B_{21} & \B_{11}
\end{array}
\right],
\label{bBinv}
\ee
and $\B^{* -1}$ is the complex conjugate of this. In Eq.~(\ref{tNt}) the 
notation $E_j \equiv E_{N_j}$ is used for the energy of the neutrino mass 
eigenstate $|N_j \rangle$, where $E_{N_j}$ is given in Eq.~(\ref{pNj}).
The states $|N_j\rangle$ ($j=1,2$) are orthogonal to each other
\be
\langle N_j | N_k \rangle = \delta_{j k}.
\label{NjNk}
\ee
We note that the  $2 \times 2$ matrix $\B$ matrix, Eq.~(\ref{bB}),
although having its two lines normalized, 
is, in general, not unitary, and therefore
\be
\langle {\cal N}_1 | {\cal N}_2 \rangle =
\sum \B_{1 j} \B_{2 j}^* \not= 0,
\quad
\langle {\cal N}_1 | {\overline {\cal N}_2} \rangle =
\sum \B_{1 j} \B_{2 j} \not= 0,
\label{cN1cN2}
\ee
i.e., the states of the heavy flavor analogs, $|{\cal N}_{\alpha} \rangle$
and/or $|{\overline {\cal N}_{\beta}} \rangle$,
are, in general, not mutually orthogonal. As a consequence 
of Eqs.~(\ref{NjNk}) and (\ref{cNs}),
these flavor analogs are normalized states
\be
\langle {\cal N}_1 | {\cal N}_1 \rangle =
\langle {\cal N}_2 | {\cal N}_2 \rangle = 1
= \langle {\overline {\cal N}_1} | {\overline {\cal N}_1} \rangle
= \langle {\overline {\cal N}_2} | {\overline {\cal N}_2} \rangle.
\label{cN1cN1}
\ee
In the LNC decay $B^+ \to \mu^+ e^- \pi^+$, Fig.~\ref{FigBmuepiLNC},
the neutrino flavor state produced in the first (production) vertex is
$| {\cal N}_2 \rangle$, and the state disappearing at the second (decay)
vertex is $|{\cal N}_1 \rangle$, cf.~Fig.~\ref{FigBLNC}.
Therefore, the relevant oscillation amplitude in this decay is
$\langle {\cal N}_1 | {\cal N}_2(t) \rangle$.
\begin{figure}[htb]
\centering\includegraphics[width=100mm]{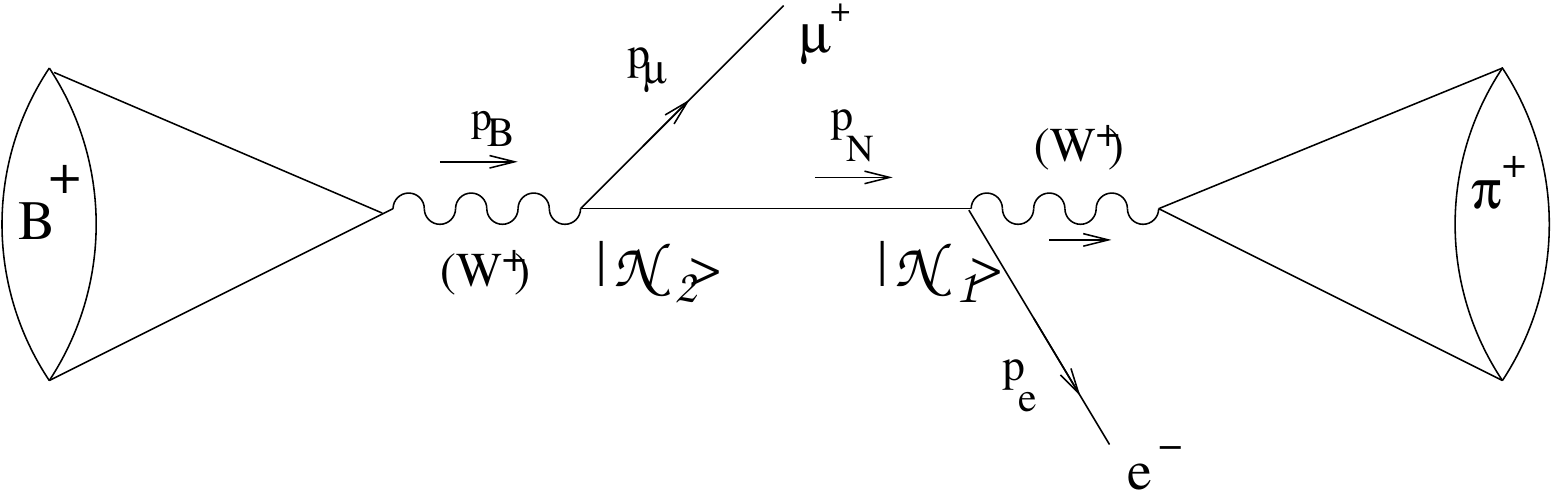}
\caption{The LNC decay $B^+ \to \mu^+ e^- \pi^+$: at the production
vertex, the $|{\cal N}_1 \rangle$ state is produced; at the decay vertex,
the $|{\cal N}_2 \rangle$ state is  absorbed.}
\label{FigBLNC}
\end{figure}

Using relations (\ref{tNta}), (\ref{cN1cN2}) and (\ref{cN1cN1}),
we obtain\footnote{
The algebra is performed in analogy with the usual 
quantum mechanics approach to light neutrino oscillations \cite{Bilenky},
except that now we have, in general, the nonorthogonality of the
two flavor states Eq.~(\ref{cN1cN2}).}
the following expression for the relevant oscillation amplitude 
$\langle {\cal N}_1 | {\cal N}_2(t) \rangle$:
\bes
\label{N1N2t}
\bea
\langle {\cal N}_1 | {\cal N}_2(t) \rangle & = &
{\big \{} \exp(- i E_{N_1} t) \B_{21}^* 
\left[ \left( \B^{* -1} \right)_{11} + 
\left( \B^{* -1} \right)_{12} \left( \B_{11} \B_{21}^* + \B_{12} \B_{22}^* \right)
\right]
\nonumber\\
&& + \exp(- i E_{N_2} t) \B_{22}^* 
\left[ \left( \B^{* -1} \right)_{21} + 
\left( \B^{* -1} \right)_{22} \left( \B_{11} \B_{21}^* + \B_{12} \B_{22}^* \right)
\right]
{\big \}}
\label{N1N2ta}
\\
& = &
\frac{1}{{\rm Det} \B^{*}} \times 
{\bigg \{} \exp(- i E_{N_1} t) \B_{21}^* 
\left[ \B_{22}^* - 
\B_{12}^* \left( \B_{11} \B_{21}^* + \B_{12} \B_{22}^* \right)
\right]
\nonumber\\
&& + \exp(- i E_{N_2} t) \B_{22}^* 
\left[ - \B_{21}^* + 
\B_{11}^* \left( \B_{11} \B_{21}^* + \B_{12} \B_{22}^* \right)
\right]
{\bigg \}}
\label{N1N2tb}
\eea
\ees
Transforming Eq.~(\ref{N1N2ta}) into Eq.~(\ref{N1N2tb}), we use for $\B^{*-1}$
the complex conjugate of identity (\ref{bBinv}).
In this quantum mechanics approach, the terms in Eq.~(\ref{N1N2t})
with $\exp(- i E_{N_j} t)$ correspond to the terms $\exp(- i p_{N_j} \cdot z)$
of the corresponding amplitude ${\cal A}(B^+ \to \mu^+ e^- \pi^+)$ in
Eq.~(\ref{calAMLNCpl}). If the two approaches are to be consistent with
each other, then the ratio of the coefficients at
$\exp(- i E_{N_1} t)$ and $\exp(- i E_{N_2} t)$ in Eq.~(\ref{N1N2tb})
is equal to the ratio of the coefficients at
$\exp(- i p_{N_1} \cdot z)$ and $\exp(- i p_{N_2} \cdot z)$
in Eq.~(\ref{calAMLNCpl}). This means that, for consistency, we
need to have\footnote{Keeping in mind that, according to Eq.~(\ref{cNs}),
we have $B_{e N_j} = K_1  \B_{1 j} \propto \B_{1 j}$ and 
$B_{\mu N_j}=K_2  \B_{2 j}  \propto B_{2 j}$.}
\bea
\frac{ \B_{22}^* - \B_{12}^* \left( \B_{11} \B_{21}^* + \B_{12} \B_{22}^* \right)}
{-\B_{21}^* + \B_{11}^*  \left( \B_{11} \B_{21}^* + \B_{12} \B_{22}^* \right)}
& = & \frac{\B_{11}}{\B_{12}}.
\label{id}
\eea
By direct cross-multiplication, 
it is straightforward to check that this identity really holds. In checking
this identity, it is enough to use only the normalization of the lines of 
the $\B$ matrix, Eq.~(\ref{bB}): $|\B_{11}|^2 + |\B_{12}|^2=1$.

In an analogous way, we can check that this quantum mechanics approach is
consistent with the amplitude approach of the main text in the other cases:
\begin{itemize}
\item
In the LNC case $B^- \to \mu^- e^+ \pi^-$: 
in the explanation above (Fig.~\ref{FigBLNC}),
the states $|{\cal N}_2 \rangle$ and $|{\cal N}_1 \rangle$ get replaced
by $|{\overline {\cal N}_2} \rangle$ and $|{\overline {\cal N}_1} \rangle$;
cf.~Eq.~(\ref{wfs}).
\item
In the LNV case $B^+ \to \mu^+ e^+ \pi^-$: 
in the explanation above (Fig.~\ref{FigBLNC}),
the state $|{\cal N}_1 \rangle$ gets replaced
by $|{\overline {\cal N}_1} \rangle$; cf.~Eq.~(\ref{wfs}).
\item
In the LNV case $B^- \to \mu^- e^- \pi^+$: 
in the explanation above (Fig.~\ref{FigBLNC}),
the state $|{\cal N}_2 \rangle$ gets replaced
by $|{\overline {\cal N}_2} \rangle$; cf.~Eq.~(\ref{wfs}).
\end{itemize}


\begin{thebibliography}{99}

\bibitem{0nubb}
  G.~Racah,
  On the symmetry of particle and antiparticle,
  Nuovo Cimento  {\bf 14}, 322 (1937);
  W.~H.~Furry,
  On transition probabilities in double beta-disintegration,
  Phys.\ Rev.\  {\bf 56}, 1184 (1939);
H.~Primakoff and S.~P.~Rosen, Double beta decay,
Rep. Prog. Phys. {\bf 22}, 121 (1959);
  Nuclear double-beta decay and a new limit on lepton nonconservation,
  Phys.\ Rev.\  {\bf 184}, 1925 (1969);
  Baryon Number And Lepton Number Conservation Laws,
  Annu.\ Rev.\ Nucl.\ Part.\ Sci.\  {\bf 31}, 145 (1981);
  J.~Schechter and J.~W.~F.~Valle,
  Neutrinoless Double beta Decay in $SU(2) x U(1)$ Theories,
  Phys.\ Rev.\ D {\bf 25}, 2951 (1982);
 M.~Doi, T.~Kotani and E.~Takasugi,
  Double beta Decay and Majorana Neutrino,
  Prog.\ Theor.\ Phys.\ Suppl.\  {\bf 83}, 1 (1985);
  S.~R.~Elliott and J.~Engel,
  Double beta decay,
  J.\ Phys.\ G  {\bf 30}, R183 (2004)
  [hep-ph/0405078];
  V.~A.~Rodin, A.~Faessler, F.~Simkovi\'c and P.~Vogel,
  Assessment of uncertainties in QRPA $0\nu\beta\beta$-decay nuclear matrix elements,
  Nucl.\ Phys.\ A {\bf 766}, 107 (2006);
  A {\bf 793}, 213(E) (2007)
  [arXiv:0706.4304 [nucl-th]].

\bibitem{scatt1}
  W.~-Y.~Keung and G.~Senjanovi\'c,
  Majorana Neutrinos And The Production Of The Right-handed Charged Gauge Boson,
  Phys.\ Rev.\ Lett.\  {\bf 50}, 1427 (1983);
  V.~Tello, M.~Nemev\v{s}ek, F.~Nesti, G.~Senjanovi\'c and F.~Vissani,
  Left-Right Symmetry: from LHC to Neutrinoless Double Beta Decay,
  Phys.\ Rev.\ Lett.\  {\bf 106}, 151801 (2011)
  [arXiv:1011.3522 [hep-ph]];
  M.~Nemev\v{s}ek, F.~Nesti, G.~Senjanovi\'c and V.~Tello,
  Neutrinoless Double Beta Decay: Low Left-Right Symmetry Scale?,
  arXiv:1112.3061 [hep-ph];
  G.~Senjanovi\'c,
  Neutrino mass: From LHC to grand unification,
  Riv.\ Nuovo Cimento  Soc. Ital. Fis. {\bf 034}, 1 (2011).

\bibitem{scatt234}
  W.~Buchm\"uller and C.~Greub,
  Heavy Majorana neutrinos in electron - positron and electron - proton collisions,
  Nucl.\ Phys.\ B {\bf 363}, 345 (1991);
  M.~Kohda, H.~Sugiyama and K.~Tsumura,
  Lepton number violation at the LHC with leptoquark and diquark,
  Phys.\ Lett.\ B {\bf 718}, 1436 (2013)
  [arXiv:1210.5622 [hep-ph]].

\bibitem{scatt3}
  J.~C.~Helo, M.~Hirsch and S.~Kovalenko,
  Heavy neutrino searches at the LHC with displaced vertices,
  Phys.\ Rev.\ D {\bf 89}, 073005 (2014)
  [arXiv:1312.2900 [hep-ph]].

\bibitem{scattDev}
  C.~Y.~Chen and P.~S.~Bhupal Dev,
  Multi-Lepton Collider Signatures of Heavy Dirac and Majorana Neutrinos,
  Phys.\ Rev.\ D {\bf 85}, 093018 (2012)
  [arXiv:1112.6419 [hep-ph]];
  C.~Y.~Chen, P.~S.~Bhupal Dev and R.~N.~Mohapatra,
  Probing Heavy-Light Neutrino Mixing in Left-Right Seesaw Models at the LHC,
  Phys.\ Rev.\ D {\bf 88}, 033014 (2013)
  [arXiv:1306.2342 [hep-ph]];
  P.~S.~Bhupal Dev, A.~Pilaftsis and U.~k.~Yang,
  New Production Mechanism for Heavy Neutrinos at the LHC,
  Phys.\ Rev.\ Lett.\  {\bf 112}, 081801 (2014)
  [arXiv:1308.2209 [hep-ph]];

\bibitem{scattDas}
  A.~Das and N.~Okada,
  Inverse seesaw neutrino signatures at the LHC and ILC,
  Phys.\ Rev.\ D {\bf 88}, 113001 (2013)
  [arXiv:1207.3734 [hep-ph]].
  A.~Das, P.~S.~Bhupal Dev and N.~Okada,
  Direct bounds on electroweak scale pseudo-Dirac neutrinos from $\sqrt s=8$ TeV LHC data,
  Phys.\ Lett.\ B {\bf 735}, 364 (2014)
  [arXiv:1405.0177 [hep-ph]].

\bibitem{RMDs}
 L.~S.~Littenberg and R.~E.~Shrock,
  Upper bounds on lepton number violating meson decays,
  Phys.\ Rev.\ Lett.\  {\bf 68}, 443 (1992);
  Implications of improved upper bounds on $|\Delta L| = 2$ processes,
  Phys.\ Lett.\ B {\bf 491}, 285 (2000)
  [hep-ph/0005285];
 C.~Dib, V.~Gribanov, S.~Kovalenko and I.~Schmidt,
  K meson neutrinoless double muon decay as a probe of neutrino masses and mixings,
  Phys.\ Lett.\ B {\bf 493}, 82 (2000)
  [hep-ph/0006277];
  A.~Ali, A.~V.~Borisov and N.~B.~Zamorin,
  Majorana neutrinos and same sign dilepton production at LHC and in rare meson decays,
  Eur.\ Phys.\ J.\ C {\bf 21}, 123 (2001)
  [hep-ph/0104123];
  M.~A.~Ivanov and S.~G.~Kovalenko,
  Hadronic structure aspects of $K^{+} \to \pi^- + l^+_1 + l^+_2$ decays,
  Phys.\ Rev.\ D {\bf 71}, 053004 (2005)
  [hep-ph/0412198];
  A.~de Gouvea and J.~Jenkins,
  Survey of lepton number violation via effective operators,
  Phys.\ Rev.\ D {\bf 77}, 013008 (2008)
  [arXiv:0708.1344 [hep-ph]];
  J.~C.~Helo, S.~Kovalenko and I.~Schmidt,
  Sterile neutrinos in lepton number and lepton flavor violating decays,
  Nucl.\ Phys.\ B {\bf 853}, 80 (2011)
  [arXiv:1005.1607 [hep-ph]];
  N.~Quintero, G.~L\'opez Castro and D.~Delepine,
  Lepton number violation in top quark and neutral B meson decays,
  Phys.\ Rev.\ D {\bf 84}, 096011 (2011)
  [Phys.\ Rev.\ D {\bf 86}, 079905 (2012)]
  [arXiv:1108.6009 [hep-ph]];
  G.~L.~Castro and N.~Quintero,
  Bounding resonant Majorana neutrinos from four-body B and D decays,
  Phys.\ Rev.\ D {\bf 87}, 077901 (2013)
  [arXiv:1302.1504 [hep-ph]];
  A.~Abada, A.~M.~Teixeira, A.~Vicente and C.~Weiland,
  Sterile neutrinos in leptonic and semileptonic decays,
  JHEP {\bf 1402}, 091 (2014)
  [arXiv:1311.2830 [hep-ph]];
  Y.~Wang, S.~S.~Bao, Z.~H.~Li, N.~Zhu and Z.~G.~Si,
  Study Majorana neutrino contribution to B-meson cemi-leptonic rare decays,
  Phys.\ Lett.\ B {\bf 736}, 428 (2014)
  [arXiv:1407.2468 [hep-ph]].

\bibitem{Atre}
  A.~Atre, T.~Han, S.~Pascoli and B.~Zhang,
  The search for heavy Majorana neutrinos,
  JHEP {\bf 0905}, 030 (2009)
  [arXiv:0901.3589 [hep-ph]], and references therein.

\bibitem{Boya} 
  D.~Boyanovsky,
  Nearly degenerate heavy sterile neutrinos in cascade decay: mixing and oscillations,
  Phys.\ Rev.\ D {\bf 90}, 105024 (2014)
  [arXiv:1409.4265 [hep-ph]].


\bibitem{CDKK}
  G.~Cveti\v{c}, C.~Dib, S.~K.~Kang and C.~S.~Kim,
  Probing Majorana neutrinos in rare $K$ and $D, ~D_s, B, B_c$ meson decays,
  Phys.\ Rev.\ D {\bf 82}, 053010 (2010)
  [arXiv:1005.4282 [hep-ph]].

\bibitem{CDK}
  G.~Cveti\v{c}, C.~Dib and C.~S.~Kim,
  Probing Majorana neutrinos in rare $\pi^+ \to e^+ e^+ \mu^- \nu$ decays,
  JHEP {\bf 1206}, 149 (2012)
  [arXiv:1203.0573 [hep-ph]].

\bibitem{CKZ}
  G.~Cveti\v{c}, C.~S.~Kim and J.~Zamora-Sa\'a,
  CP violations in $\pi^{\pm}$ meson decay,
  J.\ Phys.\ G {\bf 41}, 075004 (2014)
  [arXiv:1311.7554 [hep-ph]].


\bibitem{CKZ2}
  G.~Cveti\v{c}, C.~S.~Kim and J.~Zamora-Sa\'a,
  CP violation in lepton number violating semihadronic decays of $K,D,D_s,B,B_c$,
  Phys.\ Rev.\ D {\bf 89}, 093012 (2014)
  [arXiv:1403.2555 [hep-ph]].

\bibitem{DCK}
  C.~O.~Dib, M.~Campos and C.~S.~Kim,
  CP violation with Majorana neutrinos in $K$ meson decays,
  JHEP {\bf 1502}, 108 (2015)
  [arXiv:1403.8009 [hep-ph]].


\bibitem{CDKZ}
  G.~Cveti\v{c}, C.~Dib, C.~S.~Kim and J.~Zamora-Sa\'a,
  Probing the Majorana neutrinos and their CP violation in decays of charged scalar mesons $\pi, K, D, D_s, B, B_c$,
Symmetry {\bf 7}, 726 (2015)
  [arXiv:1503.01358 [hep-ph]].

\bibitem{Pontecorvo}
  B.~Pontecorvo,
  Inverse beta processes and nonconservation of lepton charge,
  Zh.\ Eksp.\ Teor.\ Fiz.\  {\bf 34}, 247 (1957) [Sov.\ Phys.\ JETP {\bf 7}, 172 (1958)];
  Neutrino experiments and the problem of conservation of leptonic charge,
 Zh.\ Eksp.\ Teor.\ Fiz.\  {\bf 53}, 1717 (1967)  [Sov.\ Phys.\ JETP {\bf 26}, 984 (1968)].

\bibitem{oscatm}
  Y.~Fukuda {\it et al.}  [Super-Kamiokande Collaboration],
  Evidence for oscillation of atmospheric neutrinos,
  Phys.\ Rev.\ Lett.\  {\bf 81}, 1562 (1998)
  [hep-ex/9807003].

\bibitem{oscsol}
  Q.~R.~Ahmad {\it et al.}  [SNO Collaboration],
  Direct evidence for neutrino flavor transformation from neutral current interactions in the Sudbury Neutrino Observatory,
Phys.\ Rev.\ Lett.\  {\bf 89}, 011301 (2002) [nucl-ex/0204008];
  P.~Lipari,
  CP violation effects and high-energy neutrinos,
  Phys.\ Rev.\ D {\bf 64}, 033002 (2001)
  [hep-ph/0102046];
  Z.~Rahman, A.~Dasgupta and R.~Adhikari,
  Discovery reach of CP violation in neutrino oscillation experiments with standard and non-standard interactions,
  arXiv:1210.2603 [hep-ph].
  Which baseline for neutrino factory could be better for discovering CP violation in neutrino oscillation for standard and non-standard interactions?,
  arXiv:1210.4801 [hep-ph].

\bibitem{oscnuc}
  K.~Eguchi {\it et al.}  [KamLAND Collaboration],
  First results from KamLAND: Evidence for reactor anti-neutrino disappearance,
  Phys.\ Rev.\ Lett.\  {\bf 90}, 021802 (2003)
  [hep-ex/0212021].

\bibitem{PlanckColl} 
  P.~A.~R.~Ade {\it et al.}  [Planck Collaboration],
  Planck 2013 results. XVI. Cosmological parameters,
  Astron.\ Astrophys.\  {\bf 571}, A16 (2014)
  [arXiv:1303.5076 [astro-ph.CO]].

\bibitem{seesaw}
  P.~Minkowski,
  $\mu \to e \gamma$ at a rate of one out of $10^9$ muon decays?,
  Phys.\ Lett.\ B {\bf 67}, 421 (1977);
M.~Gell-Mann, P.~Ramond and R.~Slansky, in Sanibel Conference, 
``The family group in Grand Unified Theories,'' Febr.~1979, Report No.~CALT-68-709, reprinted in hep-ph/9809459;
"Complex Spinors and Unified Theories," Print 80-0576, 
published in: D.~Freedman et al. (Eds.), {\it Supergravity}, North-Holland, Amsterdam, 1979;
  T.~Yanagida,
  Horizontal Symmetry And Masses Of Neutrinos,
  Conf.\ Proc.\ C {\bf 7902131}, 95 (1979);
S.~L.~Glashow, in: M.~Levy et al. (Eds.), {\it Quarks and Leptons}, Cargese,
Plenum, New York, 1980, p.~707;
  R.~N.~Mohapatra and G.~Senjanovi\'c,
  Neutrino mass and spontaneous parity violation,
  Phys.\ Rev.\ Lett.\  {\bf 44}, 912 (1980).

\bibitem{WWMMD}
  D.~Wyler and L.~Wolfenstein,
  Massless neutrinos in left-right symmetric models,
  Nucl.\ Phys.\ B {\bf 218}, 205 (1983);
  E.~Witten,
  Symmetry breaking patterns in superstring models,
  Nucl.\ Phys.\ B {\bf 258}, 75 (1985);
  R.~N.~Mohapatra and J.~W.~F.~Valle,
  Neutrino mass and baryon number nonconservation in superstring models,
  Phys.\ Rev.\ D {\bf 34}, 1642 (1986);
  M.~Malinsky, J.~C.~Romao and J.~W.~F.~Valle,
  Novel supersymmetric SO(10) seesaw mechanism,
  Phys.\ Rev.\ Lett.\  {\bf 95}, 161801 (2005)
  [hep-ph/0506296];
  P.~S.~Bhupal Dev and R.~N.~Mohapatra,
  TeV scale inverse seesaw in $SO(10)$ and leptonic non-unitarity ffects,
  Phys.\ Rev.\ D {\bf 81}, 013001 (2010)
  [arXiv:0910.3924 [hep-ph]];
  P.~S.~Bhupal Dev and A.~Pilaftsis,
  Minimal Radiative Neutrino Mass Mechanism for Inverse Seesaw Models,
  Phys.\ Rev.\ D {\bf 86}, 113001 (2012)
  [arXiv:1209.4051 [hep-ph]];
  C.~H.~Lee, P.~S.~Bhupal Dev and R.~N.~Mohapatra,
  Natural TeV-scale left-right seesaw mechanism for neutrinos and experimental tests,
  Phys.\ Rev.\ D {\bf 88}, 093010 (2013)
  [arXiv:1309.0774 [hep-ph]].

\bibitem{nuMSM}
  T.~Asaka, S.~Blanchet and M.~Shaposhnikov,
  The $\nu$MSM, dark matter and neutrino masses,
  Phys.\ Lett.\ B {\bf 631}, 151 (2005)
  [hep-ph/0503065];
  T.~Asaka and M.~Shaposhnikov,
  The $\nu$MSM, dark matter and baryon asymmetry of the universe,
  Phys.\ Lett.\ B {\bf 620}, 17 (2005)
  [hep-ph/0505013].

\bibitem{HeAAS}
  F.~del Aguila, J.~A.~Aguilar-Saavedra, J.~de Blas and M.~Zralek,
  Looking for signals beyond the neutrino Standard Model,
  Acta Phys.\ Polon.\ B {\bf 38}, 3339 (2007)
  [arXiv:0710.2923 [hep-ph]];
  X.~G.~He, S.~Oh, J.~Tandean and C.~C.~Wen,
  Large Mixing of Light and Heavy Neutrinos in Seesaw Models and the LHC,
  Phys.\ Rev.\ D {\bf 80}, 073012 (2009)
  [arXiv:0907.1607 [hep-ph]].

\bibitem{KS}
  J.~Kersten and A.~Y.~Smirnov,
  Right-handed neutrinos at CERN LHC and the mechanism of neutrino mass generation,
  Phys.\ Rev.\ D {\bf 76}, 073005 (2007)
  [arXiv:0705.3221 [hep-ph]].

\bibitem{AMP}
  A.~Ibarra, E.~Molinaro and S.~T.~Petcov,
  TeV scale see-saw mechanisms of neutrino mass generation, the Majorana nature of the heavy singlet neutrinos and $0\nu\beta\beta$-decay,
  JHEP {\bf 1009}, 108 (2010)
  [arXiv:1007.2378 [hep-ph]].

\bibitem{NSZ}
  M.~Nemev\v{s}ek, G.~Senjanovi\'c and Y.~Zhang,
  Warm dark matter in low scale left-right theory,
  JCAP {\bf 1207}, 006 (2012)
  [arXiv:1205.0844 [hep-ph]].

\bibitem{Pilaftsis} 
  A.~Pilaftsis,
  CP violation and baryogenesis due to heavy Majorana neutrinos,
  Phys.\ Rev.\ D {\bf 56}, 5431 (1997)
  [hep-ph/9707235];
  S.~Bray, J.~S.~Lee and A.~Pilaftsis,
  Resonant CP violation due to heavy neutrinos at the LHC,
  Nucl.\ Phys.\ B {\bf 786}, 95 (2007)
  [hep-ph/0702294 [HEP-PH]].

\bibitem{Shapo}
  D.~Gorbunov and M.~Shaposhnikov,
  How to find neutral leptons of the $\nu$MSM?,
  JHEP {\bf 0710}, 015 (2007)
  [JHEP {\bf 1311}, 101 (2013)]
  [arXiv:0705.1729 [hep-ph]];
  A.~Boyarsky, O.~Ruchayskiy and M.~Shaposhnikov,
  The role of sterile neutrinos in cosmology and astrophysics,
  Annu.\ Rev.\ Nucl.\ Part.\ Sci.\  {\bf 59}, 191 (2009)
  [arXiv:0901.0011 [hep-ph]];
  L.~Canetti, M.~Drewes and M.~Shaposhnikov,
  Sterile neutrinos as the origin of dark and baryonic matter,
  Phys.\ Rev.\ Lett.\  {\bf 110}, 061801 (2013)
  [arXiv:1204.3902 [hep-ph]];
  L.~Canetti, M.~Drewes, T.~Frossard and M.~Shaposhnikov,
  Dark matter, baryogenesis and neutrino oscillations from right handed neutrinos,
  Phys.\ Rev.\ D {\bf 87}, 093006 (2013)
  [arXiv:1208.4607 [hep-ph]].


\bibitem{lsseesaw1} 
L.~Canetti, M.~Drewes and B.~Garbrecht,
Probing leptogenesis with GeV-scale sterile neutrinos at LHCb and Belle II,
Phys.\ Rev.\ D {\bf 90}, 125005 (2014) 
[arXiv:1404.7114 [hep-ph]]. 

\bibitem{lsseesaw2}
M.~Drewes and B.~Garbrecht,
Experimental and cosmological constraints on heavy neutrinos, 
arXiv:1502.00477 [hep-ph]. 

\bibitem{Boya2} 
  D.~Boyanovsky and L.~Lello,
  Time evolution of cascade decay,
  New J.\ Phys.\  {\bf 16}, 063050 (2014)
  [arXiv:1403.6366 [hep-ph]];
  D.~Boyanovsky,
  Space–time evolution of heavy sterile neutrinos in cascade decays,
  Nucl.\ Phys.\ B {\bf 888}, 248 (2014)
  [arXiv:1406.5739 [hep-ph]].

\bibitem{CERN-SPS}
W. Bonivento {\it et al.}, Report Nos. CERN-SPSC-2013-024, CERN-EOI-010,
[arXiv:1310.1762 [hep-ex]];
R.~Jacobson, Search for heavy neutral neutrinos at the SPS,
presented at {\it High Energy Physics in the LHC Era,\/}, UTFSM,
Valpara\'{\i}so, Chile, December 16-20, 2014,
https://indico.cern.ch/event/252857/contribution/215

\bibitem{Gronau}
  M.~Gronau, C.~N.~Leung and J.~L.~Rosner,
  Extending limits on neutral heavy leptons,
  Phys.\ Rev.\ D {\bf 29}, 2539 (1984).

\bibitem{commKim}
  C.~Dib and C.~S.~Kim,
  Remarks on the lifetime of sterile neutrinos and the effect on detection of rare meson decays $M^+ \to M^{\prime}-\ell^+\ell^+$,
  Phys.\ Rev.\ D {\bf 89}, 077301 (2014)
  [arXiv:1403.1985 [hep-ph]].


\bibitem{Glash}
  A.~G.~Cohen, S.~L.~Glashow and Z.~Ligeti,
  Disentangling neutrino oscillations,
  Phys.\ Lett.\ B {\bf 678}, 191 (2009)
  [arXiv:0810.4602 [hep-ph]].

\bibitem{Bilenky}
S.~Bilenky, {\it Introduction to the Physics of Massive and Mixed Neutrinos\/},
Lecture Notes in Physics (Springer Verlag, Berlin, Heidelberg, 2010), Vol.~817.

\bibitem{Giunti}
  C.~Giunti and C.~W.~Kim,
  Quantum mechanics of neutrino oscillations,
  Found.\ Phys.\ Lett.\  {\bf 14}, 213 (2001)
  [hep-ph/0011074].

\bibitem{DBP}
  F.~F.~Deppisch, P.~S.~Bhupal Dev and A.~Pilaftsis,
  Neutrinos and collider physics,
  arXiv:1502.06541 [hep-ph].

\bibitem{charm}
  F.~Bergsma {\it et al.}  [CHARM Collaboration],
  A search for decays of heavy neutrinos in the mass range $0.5$ GeV to $2.8$ GeV,
  Phys.\ Lett.\ B {\bf 166}, 473 (1986);
  P.~Vilain {\it et al.}  [CHARM II Collaboration],
Search for heavy isosinglet neutrinos,
  Phys.\ Lett.\ B {\bf 343}, 453 (1995); {\bf 351}, 387(E) (1995);
  J.~Orloff, A.~N.~Rozanov and C.~Santoni,
  Limits on the mixing of tau neutrino to heavy neutrinos,
  Phys.\ Lett.\ B {\bf 550}, 8 (2002)
  [hep-ph/0208075].

\bibitem{NuTeV}
  A.~Vaitaitis {\it et al.}  [NuTeV and E815 Collaborations],
  Search for neutral heavy leptons in a high-energy neutrino beam,
  Phys.\ Rev.\ Lett.\  {\bf 83}, 4943 (1999)
  [hep-ex/9908011].


\bibitem{delphi}
  P.~Abreu {\it et al.}  [DELPHI Collaboration],
  Search for neutral heavy leptons produced in Z decays,
  Z.\ Phys.\ C {\bf 74}, 57 (1997); {\bf 75}, 580(E) (1997).

\bibitem{Belle}
  D.~Liventsev {\it et al.}  [Belle Collaboration],
  Search for heavy neutrinos at Belle,
  Phys.\ Rev.\ D {\bf 87}, 071102 (2013)
  [arXiv:1301.1105 [hep-ex]].


\end{thebibliography}
\end{document}